\documentclass[oupdraft]{bio}
\usepackage{fix-cm}
\usepackage{url}
\usepackage{statmath}
\usepackage{dsfont}
\usepackage{algorithm}
\usepackage{algpseudocode}
\usepackage{amsmath}
\usepackage{rotating}
\usepackage{multirow}
\usepackage{threeparttable} 
\usepackage[margin=1in]{geometry}  
\usepackage{placeins}

\numberwithin{equation}{section}
\history{}

\begin{document}

\title{
Sequential Design with Derived Win Statistics}

\author{BAOSHAN ZHANG, \quad YUAN WU$^\ast$\\[4pt]
\textit{Department of Biostatistics and Bioinformatics, School of Medicine, Duke University}
\\[2pt]
{yuan.wu@duke.edu}}

\markboth%
{B. Zhang and Y. Wu}
{Sequential design based on Derived Win Statistics}

\maketitle

\footnotetext{Corresponding Address: 2424 Erwin Rd, Durham, NC 27705, USA}

\begin{abstract}
{The Win Ratio has gained significant traction in cardiovascular trials as a novel method for analyzing composite endpoints \citep{pocock2012win}. Compared with conventional approaches based on time to the first event, the Win Ratio accommodates the varying priorities and types of outcomes among components, potentially offering greater statistical power by fully utilizing the information contained within each outcome. However, studies using Win Ratio have largely been confined to fixed design, limiting flexibility for early decisions, such as stopping for futility or efficacy. Our study proposes a sequential design framework incorporating multiple interim analyses based on Win Ratio or Net Benefit statistics. Moreover, we provide rigorous proof of the canonical joint distribution for sequential Win Ratio and Net Benefit statistics, and an algorithm for sample size determination is developed. We also provide results from a finite sample simulation study, which show that our proposed method controls Type I error maintains power level, and has a smaller average sample size than the fixed design. A real study of cardiovascular study is applied to illustrate the proposed method.}
{Sequential Design; Win Ratio; Net Benefit; U-Statistics; Hierarchical Composite Endpoint}
\end{abstract}

\newpage
\section{Introduction}
\label{sec: Intro}

The use of composite endpoints as primary outcomes is common in randomized clinical trials, particularly in cardiovascular (CV) trials. A frequently used composite endpoint in CV trials is Major Adverse Cardiac Events, which typically include CV death, myocardial infarction, stroke, hospitalization, and other related events \citep{sharma2020impact}. The popularity of composite endpoints stems from two key advantages: (1) Potentially increase the event rate and statistical power, thereby reducing the required sample size; (2) Eliminate the need for multiplicity adjustments, and (3) Provide a unified measure of treatment effect.

There are two primary approaches to comparing composite endpoints between subjects: by event time or by clinical priority. The former approach, often called the traditional composite endpoint (TCE), is defined as the time to the first event \citep{pocock1997clinical}. The second approach, known as the hierarchical composite endpoint (HCE), involves comparing each component of the endpoint between the subjects based on a predefined clinical priority order of HCE \citep{buyse2010generalized, pocock2012win}. Compared to TCE, HCE addresses the challenges of analyzing prioritized outcomes by allowing different components of the composite endpoint to carry varying levels of clinical importance. In addition, non-event outcomes with clinical meaning could also be included in HCE. A more detailed discussion of the advantages and challenges of HCE in comparison to TCE can be found in \citeauthor{redfors2020win} (\citeyear{redfors2020win}).

Given the advantages of HCE, \cite{buyse2010generalized} introduced the concept of Net Benefit (NB) based on HCE, defined as the difference between the probability that a treated subject wins against a control and the probability that a control wins against a treated subject, to assess treatment effects. \citet{pocock2012win} later proposed the Win Ratio (WR) as a metric for quantifying treatment effects on composite endpoints by the ratio between win and loss proportion for pairwise comparison, building on a similar hierarchical approach. The large-sample theory for the WR in fixed design studies has been rigorously developed using U-statistics \citep{bebu2016large}. To facilitate hypothesis testing and study design, \citet{luo2017weighted} provided a closed-form variance estimator under the null hypothesis. The asymptotic properties of WR can be readily extended to the NB and other test statistics derived from win statistics.
Additionally, \citet{mao2022sample} introduced a sample size calculation method for fixed design studies, incorporating a combination of copula models \citep{oakes1989bivariate}, U-statistics, and numerical computations. Currently, according to the online registry ClinicalTrials.gov's data, there is a sharp rise in both the number of trials and the patients involved to adopte this Win Ratio or related methodology \cite{mao2024defining}.

To date, research has predominantly focused on fixed design methodologies. However, clinical trials often benefit from more flexible approaches that allow for early decisions based on interim data. Group sequential designs are a common type of adaptive clinical trial design that offers flexibility over fixed designs, enabling decisions regarding early stopping for futility or efficacy \citep{pocock1977group}. Group sequential tests for continuous, binary, and time-to-event outcomes are well-documented in the literature. Despite the development of fixed design methods for win statistics, sequential designs remain under-explored in the literature. \citet{bergemann4450923group} conducted a pilot study that demonstrated independent increments for sequential NB and approximately independent increments for sequential WR, both of which are crucial for developing sequential designs \citep{jennison:1997, scharfstein:1997}. Establishing the asymptotic canonical distribution is a critical component in developing sequential designs \citep{jennison:2000}. However, the canonical joint normality of sequential test statistics has yet to be formally established \citep{jennison:2000}. Up to now, there is also a lack of power analysis and corresponding sample size determination methods for sequential designs based on win statistics.

In our study, we developed a sequential design based on NB and WR test statistics. Firstly, we reviewed the fixed design based on NB and WR test statistics and its asymptotic normality in Section \ref{Sec: FixedWinStat}. Next, in Section \ref{Sec: SeqWinStat}, we extended to group sequential setting and rigorously proofed the canonical joint distribution for sequential NB and WR test statistics. Variance estimation methods are also provided in this section. Then, power analysis and type-I error control are discussed in detail in Section \ref{Sec: PowerANA}. In addition, we provide a sample size determination algorithm in Section \ref{Sec: PowerANA} considering the practical application. Section \ref{Sec: NumericalAna} conducted a detailed numerical analysis to mimic the scenario of a CV trial using HCE. Then, we illustrate our developed design based on sequential derived win statistics to a real-world example from the HF-ACTION in section \ref{Sec: RealANA}. Finally, the conclusion and discussion for this provided sequential design are provided in Section \ref{Sec: Conclusion}.

\section{Derived Win Statistics for Fixed Design}\label{Sec: FixedWinStat} 

In this section, we begin by illustrating the framework of the Win Statistics within a fixed design, particularly focusing on HCE that involve both terminal and non-terminal events. Next,  we extend the proposed framework to accommodate general mixed-type HCE. In the final part of this section, we present the asymptotic properties of the derived Win Statistics, including the Win Ratio and Net Benefit.

\subsection{Basic Setting and Consideration with Survival Outcome Only}
Firstly, we consider the HCE to consist of a terminal and a non-terminal event. The terminal event, such as the event of death, censors the event of the non-terminal event. That is to say, we could not observe the non-terminal event after the terminal event happens. This setting of the composite endpoint is commonly used in CV trials where for instance CV death is the terminal event and the non-fatal stroke, hospitalization of heart failure, or nonfatal MI is the non-terminal event as considered  \citep{luo2017weighted, james2024dapagliflozin}. To help for illustration, in this subsection, we denote the composite endpoint as the time to death $T_1$ for the terminal event and the time to hospitalization of heart failure $T_2$ as the non-terminal event. These two variables are usually correlated. In addition, $T_1$ may right-censor $T_2$ but not vice versa. As for censoring distribution, we have the following two assumptions:

\begin{Assumption}[\textit{Common Censoring}] \label{Asmp: ComCen}
We assume the censoring time $C$ is shared by all $2$-composite time-to-event endpoints $(T_1, T_2)$ within the same subject.
\end{Assumption}
In addition, following \cite{fine1999proportional}, we assume the independence of censoring time and both $2-$composite time-to-event endpoints:
\begin{Assumption}[\textit{Independence of Censoring Time and Composite Endpoints}]
We assume 
\[
C \perp T_1 \quad \text{AND} \quad C\perp T_2
\]
\end{Assumption}
\begin{remark}
    Both common censoring and independence assumption of the censoring time $C$ based on 2-composite time-to-event endpoints above could be extended to a more general case with $q-$composite  time-to-event endpoints. 
\end{remark}

Due to the censoring time $C$ and the terminal event $T_1$, we observe $Y_2 := T_2 \wedge T_1 \wedge C$ and $Y_1 := T_1 \wedge C$, along with the event indicators $\delta_2 = \mathds{1}\{Y_2 = T_2\}$ and $\delta_1 = \mathds{1}\{Y_1 = T_1\}$. Here and in the following, we define $a \wedge b$ as the minimum value between $a$ and $b$. It is important to note that if both events are non-terminal, the observation value of each non-terminal event is the minimum of the censoring time and its respective time-to-event, without involving the terminal event in taking the minimum. Suppose in a fixed random trial design, there are $m$ patients in the treatment arm and $n$ patients in the control arm. We denote observed outcome $\mathcal{D}_i= \{D_{1i}, D_{2i}\}=\{(Y_{1i}, \delta_{1i}), (Y_{2i}, \delta_{2i})\}$ for subject $i=1, \ldots, m$ in the treatment arm and $\mathcal{D'}_j = \{D_{1i}', D_{2i}'\}=\{(Y'_{1j}, \delta'_{1j}), (Y'_{2j}, \delta'_{2j})\}$ for subject $j=1, \ldots, n$ in the control arm. Here and in the sequel, we denote subject $i$ and without superscript to denote the subject in the treatment group and $j$ and superscript as in the control group. 

Furthermore, following the definition from \cite{luo2017weighted}, for two subjects $i$ and $j$, we define win indicator ($i$ over $j$) based on event of death and event of hospitalization as $W_1^{ij}=\delta_{1j}\mathds{1}\{Y_{1i}>Y'_{1j}\}$ and $W_2^{ij} = \delta_{2j}\mathds{1}\{Y_{2i}>Y'_{2j}\},$ respectively. Similarly, the loss indicator ($i$ against $j$) are $L_1^{ij}=\delta_{1i}\mathds{1}\{Y_{1i}<Y'_{1j}\}$ and $L_2^{ij} = \delta_{2i}\mathds{1}\{Y_{2i}<Y'_{2j}\},$ respectively. Then, other scenarios are inconclusive and we define tie indicators are $\Omega_1^{ij}=(1-W_1^{ij})(1-L_1^{ij})$ and $\Omega_2^{ij}=(1-W_2^{ij})(1-L_2^{ij}),$ respectively.

Next, following \citet{bebu2016large} we define the winner, loser, and tier considering the composite endpoints as $T_1, T_2$. Notice that elements in $\mathcal{D}$ and $\mathcal{D}'$ are ordered according to priority, starting with the terminal event $T_1$. Then, following the partial ordering \citep{rauch2014opportunities}, we define comparison between $\mathcal{D}_i$ and $\mathcal{D}_j'$ as 
\begin{equation}\label{eq: DefofWin}
    \begin{aligned}
        \mathcal{D}_i \succ \mathcal{D}_j', \quad &\text{if} \quad W_1^{ij}=1\\
        & \text{else if} \quad\Omega_1^{ij} \times W_2^{ij}=1.
    \end{aligned}
\end{equation} 
If $\mathcal{D}_i \succ \mathcal{D}_j'$, then $\mathcal{D}_i$ is called a winner, while $\mathcal{D}_j'$ is a loser. Similarly, we could define $\mathcal{D}_i \prec \mathcal{D}_j'$ and here $\mathcal{D}_i$ is a loser but $\mathcal{D}_j'$ is a winner. Lastly, if neither  $\mathcal{D}_i \succ \mathcal{D}_j'$ or  $\mathcal{D}_i \prec \mathcal{D}_j'$, we define this comparison as inconclusive or tie, denoted as $\mathcal{D}_i \bowtie \mathcal{D}_j'.$ By definition of (\ref{eq: DefofWin}), we show its rationality in terms of logistics. Due to the priority of $T_1$,  we first compare the $T_1$ and see if there is a winner. We could also conclude that the winner ($i$ over $j$) happens on $T_1$ if and only if $\delta_{1j}=1$. If we cannot tell the winner or the loser based on $T_1$, where $\Omega_1^{ij}=1$,  we then check the next endpoint, $T_2$. It should be noted that the indicator function here differs from the study by \cite{dong2020inverse}, as they consider $(1-W_1^{ij}) \times W_2^{ij}=1$ as the indicator of winner happens on $T_2$, which overestimates the probability of winner.   Additionally, we find that $\mathcal{D}_i\succ \mathcal{D}_j'$ only happens when death or hospitalization from the control arm is observed. Similar logistics happen in  $\mathcal{D}_i \prec \mathcal{D}_j'$.

\subsection{Win Statistics based on Mixed-Type Data Contexts}
In this subsection, we extend the data type of outcomes in the hierarchy composite endpoint to the general setting. In this case, some non-event outcomes can be included, such as quality of life score and physiological measures. 

Assume that the HCEs consist of a total of $Q$ outcomes, which may differ in data type. Following the notation above, the observed HCEs for subject $i$ in the treatment group are denoted as $\mathcal{D}_i = \{D_{1i}, D_{2i}, \ldots, D_{Qi}\}$, where the outcomes in $\mathcal{D}_i$ are ordered according to clinical priority. Furthermore, the $q$-th outcome $D_q$ in $\mathcal{D}_i$ can be continuous, binary, ordinal, or time-to-event. When $D_q$ is time-to-event, it is denoted as $D_{qi} = (Y_{qi}, \delta_{qi})$, where $Y_{qi}$ represents the observed event time and $\delta_{qi}$ is the censoring indicator. Similarly, we define observed hierarchy composite endpoints of subject $j$ in the control group are noted as $\mathcal{D}'_i= \{D'_{1i}, D'_{2i},\ldots, D'_{Qi}\}$.

Next, a pairwise comparison is performed between subject $i$ in the treatment group and subject $j$ in the control group, starting with the first outcome $D_1$ and continuing until a winner is determined. The winner, loser, or tie for each outcome $q$ between these two subjects is defined based on the data type of the outcome, as shown in Table \ref{tab:DataType}.

\begin{table}[H]
    \centering
    \begin{tabular}{||c|c|c|c||}
        \hline \hline \hline
        \textbf{Outcome $q$ Data Type} & \textbf{Winner} $W_q^{ij}$ & \textbf{Loser} $L_q^{ij}$ & \textbf{Tie} $\Omega_q^{ij}$ \\
        \hline \hline
        Binary & $\mathds{1}\{D_{qi} > D'_{qj}\}$ & $\mathds{1}\{D_{qi} < D'_{qj}\}$ & $\mathds{1}\{D_{qi} = D'_{qj}\}$ \\ 
        Continuous or Ordinal & $\mathds{1}\{D_{qi} > D'_{qj} + \epsilon_q\}$ & $\mathds{1}\{D_{qi} < D'_{qj} - \epsilon_q\}$ & $\mathds{1}\{\mid D_{qi} - D'_{qj} \mid \leq \epsilon_q\}$ \\
        Time-to-event & $\delta'_{qj} \mathds{1}\{Y_{qi} > Y_{qj}'\}$ & $\delta_{qi} \mathds{1}\{Y_{qi} < Y_{qj}'\}$ & $(1 - W_q^{ij})(1 - L_q^{ij})$ \\
        \hline \hline \hline
    \end{tabular}
    
    ~ 
    
    \caption{Pairwise Comparison Results by Data Type. $\epsilon_q$ denotes a small non-negative value representing the clinically meaningful margin for continuous or ordinal data comparisons for outcome $q$.}\label{tab:DataType}
\end{table}

In the following subsection, we proposed the kernel function and  Win- and Loss- statistics for fixed design following the logistics above. First, for each pairwise comparison between $i$ and $j$,  win kernel function $\varphi_w\left(\mathcal{D}_i; \mathcal{D}'_j\right)$ and loss kernel function $\varphi_l\left(\mathcal{D}_i; \mathcal{D}'_j\right)$ are defined as
\begin{equation}
    \varphi_w\left(\mathcal{D}_i; \mathcal{D}'_j\right)= 
    W_1^{ij}  + \sum_{q=2}^Q\left(\prod_{p=1}^{q-1}\Omega_p^{ij}\times W_{q}^{ij}\right), \quad 
    \varphi_l\left(\mathcal{D}_i; \mathcal{D}'_j\right)= 
    L_1^{ij}  + \sum_{q=2}^Q\left(\prod_{p=1}^{q-1}\Omega_p^{ij}\times L_{q}^{ij}\right).
\end{equation}
With the defined kernel function above, the Win Statistics $U_w$ and Loss Statistics $U_l$ are proposed as the summation across all $m\times n$ pairwise comparisons, where
\begin{equation}\label{eq: U_w}
    \begin{aligned}
    U_\nu &=\frac{1}{mn}\sum_{i=1}^m\sum_{j=1}^n\varphi_\nu^{ij} = \frac{1}{mn}\sum_{i=1}^m\sum_{j=1}^n\varphi_\nu\left(\mathcal{D}_i; \mathcal{D}'_j\right), \quad \nu \in\{w, l\}.\\
    \end{aligned}  
\end{equation} 
Here, $U_w$ and $U_l$ are the total number of wins and losses in the treatment group among pairwise comparisons, respectively.
For $u = w, l$, the variance $\mathbf{V}(U_u)$ can be expressed as:
\begin{align}\label{eq: varU_u}
    \mathbf{V}(U_u) = \frac{n-1}{mn}\xi_{10}^{uu} + \frac{m-1}{mn}\xi_{01}^{uu} + \frac{1}{mn}\xi_{11}^{uu},
\end{align}
where, for any $u, v = w, l$, we define
$$
\xi^{uv}_{10} = \mathbf{Cov}\left[\varphi_u\left(\mathcal{D}_1; \mathcal{D}'_1\right), \varphi_v\left(\mathcal{D}_1; \mathcal{D}'_2\right)\right], \quad
\xi^{uv}_{01} = \mathbf{Cov}\left[\varphi_u\left(\mathcal{D}_1; \mathcal{D}'_1\right), \varphi_v\left(\mathcal{D}_2; \mathcal{D}'_1\right)\right],
$$
and
$$
\xi^{uv}_{11} = \mathbf{Cov}\left[\varphi_u\left(\mathcal{D}_1; \mathcal{D}'_1\right), \varphi_v\left(\mathcal{D}_1; \mathcal{D}'_1\right)\right].
$$

\subsection{Asymptotic Property of Derived Win Statistics:  Net Benefit and Win Ratio}

Derived from the proposed Win and Loss Statistics, two additional derived Win Statistics are introduced: Net Benefit and Win Ratio. The Net Benefit, also known as the Win Difference or Win-Loss \cite{buyse2010generalized, luo2017weighted}, is defined as
$
\Delta U := U_w - U_l
$
with its corresponding estimand given by
$
\Delta \tau := \tau_w - \tau_l.
$
Furthermore, we can derive the variance of the Net Benefit (NB) as
\begin{equation}\label{eq: VarDiffU}
\begin{aligned}
    \mathbf{V}(\Delta U) 
    &= \frac{n-1}{mn}\left(\xi^{10}_{ww} + \xi^{10}_{ll}-2\xi^{10}_{wl}\right) + \frac{m-1}{mn}\left(\xi^{01}_{ww} + \xi^{01}_{ll}-2\xi^{01}_{wl}\right) + \frac{1}{mn}\left(\xi^{11}_{ww} + \xi^{11}_{ll}-2\xi^{11}_{wl}\right).
\end{aligned}
\end{equation}

Next, the Win Ratio (WR), is proposed as 
$
\Theta U := {U_w}/{U_l},
$
with its corresponding estimand given by $\Theta \tau = {\tau_w}/{\tau_l}$. Additionally, the approximate variance of the WR can be derived as
\begin{equation}\label{eq: VarLogWRatio}
    \begin{aligned}
     \mathbf{V} \left[\log (\Theta U)\right]\approx&\frac{\frac{n-1}{mn}\xi^{10}_{ww} + \frac{m-1}{mn}\xi^{01}_{ww}+\frac{1}{mn}\xi^{11}_{ww}}{\tau_w^2} + \frac{\frac{n-1}{mn}\xi^{10}_{ll} + \frac{m-1}{mn}\xi^{01}_{ll}+\frac{1}{mn}\xi^{11}_{ll}}{\tau_l^2} \\
     &- 2\frac{\frac{n-1}{mn}\xi^{10}_{wl} + \frac{m-1}{mn}\xi^{01}_{wl} + \frac{1}{mn}\xi^{11}_{wl}}{\tau_w\tau_l}.
    \end{aligned}
\end{equation}
Details of the derivation process of $\mathbf{V} \left[\log (\Theta U)\right]$ are shown in the Appendix.

Then we apply these derived Win statistics to a fixed design scenario. Let $N =m+n$ be the total sample size and suppose sample size $m$ and $n$ both tend to infinity in such a way that 
$$\frac{m}{N}\rightarrow \eta, \quad \frac{n}{N}\rightarrow {1-\eta}, \quad 0\le \eta \le 1.$$ Suppose $\xi^{10}_{uv}, \xi^{01}_{uv},$ and $ \xi^{11}_{uv}$ all have finite values for $u,v = w,l$. The joint distribution of Win Statistics \( U_w \) and \( U_l \) is asymptotically normal, which follows
\begin{equation} \label{eq: U_joint_asym}
    \sqrt{N} 
\begin{pmatrix}
U_w -\tau_w  \\
U_l -\tau_l 
\end{pmatrix}
\indist \mathcal{N}
\left(
\begin{pmatrix}
0 \\
0
\end{pmatrix},
\begin{pmatrix}
\sigma^2_{ww} & \sigma^2_{wl} \\
\sigma^2_{wl} & \sigma^2_{ll}
\end{pmatrix}
\right),
\end{equation}
where the components of the variance-covariance matrix are given by
\begin{align*}
  \sigma^2_{uv} = \frac{1}{\eta} \xi_{uv}^{10} + \frac{1}{1-\eta} \xi_{uv}^{01}, \quad u, v \in\{w, l\}.
\end{align*}
Many previous studies have shown the above results of the asymptotic joint distribution of generalized U statistics, like \cite{lehmann1999elements}, \cite{bebu2016large}, and \cite{lee2019u}. Based on the asymptotic joint \( U_w \) and \( U_l \) shown in formula (\ref{eq: U_joint_asym}), we can establish the asymptotic joint distribution of different derived versions of Win-statistics, such as the Net Benefit and Win Ratio. For example, the asymptotic joint distribution of NB, $\Delta U:= U_w - U_l$, can be established by linear transformation:  
\begin{equation}\label{eq: WinDiff}
    \sqrt{N}\left(\Delta U-\Delta \tau\right) \indist \mathcal{N} \left(0, \sigma_{ww}^2+\sigma_{ll}^2-2\sigma^2_{wl} \right),
 \end{equation}
where $\Delta \tau = \tau_w-\tau_l$.Thus, we have
\begin{equation}
    D := \frac{\Delta U-\Delta\tau}{\sqrt{\mathbf{V} {(\Delta} U)}} \indist \mathcal{N} \left(0, 1\right).
\end{equation}
It is worth stating that the results of asymptotic normality for $\Delta U$ can also be derived by the Central limit theorem.  In addition, based on the variance formula of NB ($\ref{eq: VarDiffU}$), we can also show $\mathbf{V} (\sqrt{N}\Delta U)\rightarrow \sigma_{ww}^2+\sigma_{ll}^2-2\sigma^2_{wl}$ as $N \rightarrow \infty$.

Moreover, by Delta method applying on formula (\ref{eq: U_joint_asym}), the asymptotic property of log WR, $\Theta U:=\log(U_w / U_l)$, and $\Theta \tau:= \tau_w/\tau_l$  follows
\begin{equation}\label{eq: logWRatio}
    \sqrt{N}\big(\log\left(\Theta U\right)- \log\left(\Theta \tau\right)\big) \indist \mathcal{N}\left( 0, \frac{\sigma^2_{ww}}{\tau^2_w} + \frac{\sigma^2_{ll}}{\tau^2_l}-2\frac{\sigma_{wl}^2}{\tau_w\tau_l}\right),
\end{equation}
and for variance formula (\ref{eq: VarLogWRatio}), we can show $\mathbf{V} \left[\sqrt{N}\log{(\Theta U)}\right] \rightarrow {\sigma^2_{ww}}/{\tau^2_w} + {\sigma^2_{ll}}/{\tau^2_l}-2{\sigma_{wl}^2}/{(\tau_w\tau_l)}$ as $N\rightarrow \infty$. Thus we have,
\begin{equation}
 R= \frac{\log\left(\Theta U\right)- \log\left(\Theta \tau\right)}{\sqrt{\mathbf{V} \left[\log{(\Theta U)}\right]}} \indist \mathcal{N} (0,1)
\end{equation}

Notice that both asymptotic results ( formula \ref{eq: WinDiff} and \ref{eq: logWRatio}) are tiny different from \citet{bebu2016large} and \citet{bergemann4450923group}. 

\section{Sequential Design based on Derived Win-Statistics}\label{Sec: SeqWinStat}

In this section, we consider the scenario of a sequential design with multiple interim analyses, where the test statistic in each interim is based on the Win statistics introduced in Section \ref{Sec: FixedWinStat}. This section also helps to establish the following sequential design based on the NB and WR. 

In the context of a group sequential design characterized by a total of $K$ interim analyses, denote the cumulative sample size at interim $k$ by $m_k, n_k$ for the treatment arm and the control arm, respectively, with $k=1,2,\ldots, K$. We developed the following mild condition C1 following mild conditions on the sample sizes at different monitoring times for both groups is necessary for the following theoretical development.
\begin{description}
\item
Condition C1:
\textit{Let $N=m_K+n_K$ be the total sample size and $m_0=n_0=0$. It is true that $\frac{m_k-m_{k-1}}{N}\to \eta_k$ and $\frac{n_k-n_{k-1}}{N}\to\gamma_k$, as $m_k-m_{k-1}\to\infty$ and $n_k-n_{k-1}\to\infty$ for $k=1,2,\cdots,K$, where positive constants $\{\eta_k,\gamma_k\}_{k=1}^K$ satisfies that $\sum_{k=1}^K(\eta_k+\gamma_k)=1$. $  \hfill \square$}
\end{description}
\begin{remark} \label{rem: Condition}
Condition C1 requires that the sample size ratios between two arms and among different trial stages will converge as the total sample size increases. However, in practice, researchers only need to ensure that the proportion of subjects enrolled for each arm during each stage is not negligible, one simple example is to split subjects between arms and among stages evenly. $  \hfill \square$
\end{remark}

Define sequential Win $U_{wk}$ or  Loss statistics $U_{lk}$ at $k$-th interim analysis as   
\begin{equation}\label{eq: basicUk}
    \begin{aligned}
    U_{\nu k}
    =\frac{1}{m_kn_k}\sum_{i=1}^{m_k}\sum_{j=1}^{n_k}\varphi_u^{ij}
    = \frac{1}{m_kn_k}\sum_{i=1}^{m_k}\sum_{j=1}^{n_k}\varphi_\nu\left(\mathcal{D}_i; \mathcal{D}'_j\right), \quad \nu \in \{w,l\}, k\in\{1,\ldots,K\}.
    \end{aligned}  
\end{equation}

First, we show general results of the joint asymptotic normality on both basic win and loss U statistics in each stage by H\'ajek Projection. In detail, we initially establish the asymptotic normality of a random vector associated with projections at each interim analysis. Following this, we establish the desired joint asymptotic normality of sequential Win and Loss statistics, $\{U_{wk}, U_{lk}\}_{k=1}^K$. The H\'ajek projection principle, as outlined by \citet{vander:98}, specifies that projecting  $U_{\nu k} - \tau_{\nu}$ onto the Hilbert Space $\mathcal{H}$ to be $\hat{U}_{\nu k}$, where 
\begin{equation} \label{eq: SpaceH}
    \mathcal{H}=\left\{\sum_{p=1}^{n_k} a_p(\mathcal{D}_p) + \sum_{q=1}^{m_k} b_q(\mathcal{D}_q'): \forall a_p(\cdot),  b_q(\cdot) \textit{ s.t. } \mathbb{E} a_p^2\left(\mathcal{D}_p\right) < \infty, \mathbb{E} b_q^2 \big(\mathcal{D}_p'\big) < \infty\right\},
\end{equation}
and projection on such $\mathcal{H}$ will result in 
\begin{equation}\label{Eq: RawUproj}
    \begin{aligned}
    \hat{U}_{\nu k} &= \sum_{p=1}^{n_k}\mathbb{E}\left[ U_{\nu k}-\tau_{\nu} \mid \mathcal{D}_p \right]  + \sum_{q=1}^{m_k}\mathbb{E}\left[U_{\nu k} - \tau_{\nu} \mid \mathcal{D}_q' \right], \quad  \nu = w,l, k=1,\ldots, K.
    \end{aligned}
\end{equation}
with variance formula 
\begin{equation}\label{eq: varRawProjU}
    \begin{aligned}
\mathbf{V} (\hat{U}_{\nu k})  = \frac{1}{m_k}\xi^{10}_{\nu \nu} + \frac{1}{n_k}\xi^{01}_{\nu \nu},\quad  \nu = w,l, k=1,\ldots, K.
    \end{aligned}   
\end{equation}

Under condition C1, the limitation of variance formula can be derived as 
\begin{equation}\label{eq: AppVarRawProjU}
    \begin{aligned}
    \lim \mathbf{V} \left(\sqrt{N}\hat{U}_{\nu k}\right) = \Big(\sum_{i=1}^k\eta_i\Big)^{-1}\xi^{10}_{\nu \nu} + \Big(\sum_{j=1}^k\gamma_j\Big)^{-1}\xi^{01}_{\nu \nu},\quad  \nu = w,l; k=1,\ldots, K.
    \end{aligned}   
\end{equation}

\begin{lemma}\label{lem: RawProj} \textit{Joint Asymptotic Normality of Projection for Win- and Loss- Statistics}

Considering the H\'ajek projection of $U_{\nu k} - \tau_{\nu k}$ onto $\mathcal{H}$ as $\hat{U}_{\nu k}$ defined by formula (\ref{Eq: RawUproj}), define 
$\widehat{Z}_{\nu k}:=\frac{\hat{U}_{\nu k}}{\sqrt{\mathbf{V} (\hat{U}_{\nu k})}}.$ Under Condition C1, we obtain
\begin{equation}
    \begin{aligned}
    \left(\widehat{Z}_{w1},\widehat{Z}_{l1},\ldots, \widehat{Z}_{wK},\widehat{Z}_{lK}\right)^\top &=\left(\frac{\hat{U}_{w1}}{\sqrt{\mathbf{V} (\hat{U}_{w1})}}, \frac{\hat{U}_{l1}}{\sqrt{\mathbf{V} (\hat{U}_{l1})}}, \ldots, \frac{\hat{U}_{wK}}{\sqrt{\mathbf{V} (\hat{U}_{wK})}}, \frac{\hat{U}_{lK}}{\sqrt{\mathbf{V} (\hat{U}_{lK})}}     
\right)^\top \\
&\stackrel{d}{\longrightarrow} \mathcal{N}\left(\bm{0},\bm{\Sigma}_{Z1}\right),
    \end{aligned}
\end{equation}
\
as $m_k-m_{k-1}\to \infty$ and $n_k-n_{k-1}\to\infty$ for $k=1,2,\cdots,K$. Here, $\mathcal{N}\left(\mathbf{0},\mathbf{\Sigma}_{Z1}\right)$ represents the $2K$-dimensional multivariate normal distribution with a zero mean vector $\mathbf{0}$ of dimension $2K\times 1$, and a $2K\times 2K$ symmetric variance-covariance matrix $\mathbf{\Sigma}_{Z1}$. Specifically, $\mathbf{\Sigma}_{Z1}$ with the diagonal elements are 1 and for off-diagonal elements,  $\forall \nu_1, \nu_2 = w, l$  and for $1\le p\leq q\le K$, and we have 
\begin{equation}
    \begin{aligned}
&\mathbf{Cov}\left(\widehat{Z}_{\nu_1 p}, \widehat{Z}_{\nu_2 q} \right) \\
&\rightarrow\frac{\left(\sum_{k=1}^q\eta_k\right)^{-1}\xi^{10}_{\nu_1\nu_2} + \left(\sum_{k=1}^q\gamma_k\right)^{-1}\xi^{01}_{\nu_1\nu_2}}{\sqrt{ \left[\left(\sum_{k=1}^p\eta_k\right)^{-1}\xi^{10}_{\nu_1\nu_1} + \left(\sum_{k=1}^p\gamma_k\right)^{-1}\xi^{01}_{\nu_1\nu_1}\right]\times\left[\left(\sum_{k=1}^q\eta_k\right)^{-1}\xi^{10}_{\nu_2\nu_2} + \left(\sum_{k=1}^q\gamma_k\right)^{-1}\xi^{01}_{\nu_2\nu_2}\right]}},
    \end{aligned}
\end{equation}
as $m_k-m_{k-1}\to \infty$ and $n_k-n_{k-1}\to\infty$ for $k=1,2,\cdots,K$.
$ \hfill\square$
\end{lemma}

\begin{remark}
As for the asymptotic covariance of two H\'ajek projections $Z$ statistics with the same kernel function (i.e $\nu_1=\nu_2= \nu$) but at different stages $1 \leq p < q \leq K$, we have 
$$
\mathbf{Cov}\left[\frac{\hat{U}_{\nu p}}{\sqrt{\mathbf{V} (\hat{U}_{\nu p})}}, \frac{\hat{U}_{\nu q}}{\sqrt{\mathbf{V} (\hat{U}_{\nu q})}}\right]
\rightarrow 
\sqrt{\frac{\left(\sum_{k=1}^q\eta_k\right)^{-1}\xi^{10}_{\nu\nu} + \left(\sum_{k=1}^q\gamma_k\right)^{-1}\xi^{01}_{\nu\nu}}{\left(\sum_{k=1}^p\eta_k\right)^{-1}\xi^{10}_{\nu\nu} + \left(\sum_{k=1}^p\gamma_k\right)^{-1}\xi^{01}_{\nu\nu}}},
$$
which is also the limit of $\sqrt{\mathbf{V} (\hat{U}_{\nu q})/\mathbf{V} (\hat{U}_{\nu p})}$. The above results are the independent increments of the sequential statistics by \citet{kim2020independent}. Independent increments only happen at two basic win Statistics with the same kernel function. Detailed proof of lemma \ref{lem: RawProj} is shown in Appendix.$ \hfill \square$
\end{remark}

Based on the results of Lemma \ref{lem: RawProj}, we can also derive the asymptotic distribution of non-normalized H\'ajek sequential basic statistics. Specifically, we have
\begin{equation}\label{eq: HajekUnnormalized}
    \sqrt{N}\left(\hat{U}_{w1}, \hat{U}_{l1}, \ldots, \hat{U}_{wK}, \hat{U}_{lK}\right)^\top \indist \mathcal{N}\left(\bm{0}, \bm{\Sigma}_{Z2}\right),
\end{equation}
where for the asymptotic variance-covariance matrix $\bm{\Sigma}_{Z2}$, $\forall \nu_1, \nu_2 \in\{ w, l\}$  and for $1\le p\leq q\le K$, and we have 
\[
\lim \mathbf{Cov}\left(\sqrt{N} \hat{U}_{\nu_1p}, \sqrt{N}\hat{U}_{\nu_2q}\right) =  \Big(\sum_{k=1}^q\eta_k\Big)^{-1}\xi^{10}_{\nu_1 \nu_2} + \Big(\sum_{k=1}^q\gamma_k\Big)^{-1}\xi^{01}_{\nu_1 \nu_2}.
\]

Based on Lemma \ref{lem: RawProj}, we can establish the  joint asymptotic normality of sequential standardized  Win statistics (seq-SW) and Loss statistics (seq-SL) $\{Z_{wk}, Z_{lk}\}, k=1,\ldots, K$, where  $$Z_{\nu k} = \frac{U_{\nu k}-\tau_\nu}{\sqrt{\mathbf{V} (U_{\nu k})}}, \quad \nu = w,l.$$

\begin{theorem} 
\textit{Joint Asymptotic Normality for Sequential Win and Loss Statistics}
\label{thm: rawWin}
\begin{enumerate} 
   \item \textit{(Standardized Version)} 
   Considering the joint of seq-SWs and seq-SLs $\{Z_{wk}, Z_{lk}\}, k=1,\ldots, K$, under Condition C1, we have
   \begin{equation}
            \begin{aligned}
        \Big(Z_{w1}, Z_{l1}, \ldots, Z_{wK}, Z_{lK}\Big)^\top &:=  \left(\frac{{U}_{w1}-\tau_w}{\sqrt{\mathbf{V} ({U}_{w1})}}, \frac{{U}_{l1}-\tau_l}{\sqrt{\mathbf{V} ({U}_{l1})}}, \ldots, \frac{{U}_{wK}-\tau_w}{\sqrt{\mathbf{V} ({U}_{wK})}}, \frac{{U}_{lK}-\tau_l}{\sqrt{\mathbf{V} ({U}_{lK})}} \right)^\top \\
        &\indist\mathcal{N} \left(\mathbf{0},\mathbf{\Sigma}_{Z1}\right),
        \end{aligned}
   \end{equation}
    as $m_k-m_{k-1}\to\infty$ and $n_k-n_{k-1}\to\infty$ for $k=1,2,\cdots,K$, 
    where $\mathbf{V} (U_{\nu k}), \nu = w,l, k=1,\ldots,K$ can be derived based on formula (\ref{eq: varU_u}) by substituting $m_k, n_k$ for $m,n$. And variance-covariance matrix $\mathbf{\Sigma}_{Z1}$ are evaluated in lemma \ref{lem: RawProj}. 
    \item \textit{(Non-Standardized Version)} For the sequential non-standardized Win- (seq-NSWs) and Loss- statistics (seq-NSLs) $\{U_{wk}, U_{lk}\}, k=1,\ldots, K$, under Condition C1, we have
    \begin{equation}\label{eq: KstageRawJoint}
    \sqrt{N}\left(U_{w1}-\tau_{w}, U_{l1}-\tau_{l}, \ldots, U_{wK}-\tau_{w}, U_{lK}-\tau_{l}\right) \indist \mathcal{N}\left(\mathbf{0}, \mathbf{\Sigma}_{Z2}\right),
    \end{equation}
    where $\mathbf{\Sigma}_{Z2}$ is evaluated in formula (\ref{eq: HajekUnnormalized}).
    \end{enumerate}
$\hfill\square$
\end{theorem} 
\begin{remark}
It is clear that the non-standardized version of Theorem \ref{thm: rawWin} is an extension to the K-stage sequential design from the fixed design of the asymptotic joint normal distribution of  Win- and Loss-statistics from \cite{bebu2016large} (formula \ref{eq: U_joint_asym}). Moreover, this theorem is vital to help establish the following asymptotic normality results of our sequential derived Win- and Statistics (e.g., NB and WR).
Detailed proof of Theorem \ref{thm: rawWin} is shown in Appendix \ref{app: Theorem1}. Part 2 of the non-standardized version can be derived from the standardized version in Part 1.
$\hfill \square$
\end{remark}

With the results of  Theorem \ref{thm: rawWin}, we construct logically the derived Win statistics and find their joint asymptotic distribution. For each interim with $U_{uk}$, marginal asymptotic property of Win Statistics derived from  $U_{uk}$ in Section \ref{Sec: Fixed_IPCWWinStat} can be applied to $U_{uk}$. That is to say, for each interim $k$, we have Net Benefit $\Delta U_k$ and Win Ratio $\Theta U_k$ which are both derived from Win- and Loss- statistics  $\{Z_{wk}, Z_{lk}\}, k=1,\ldots, K$ and their standardized versions are defined as
\begin{equation}
    D_k:=\frac{\Delta U_k-\Delta\tau}{\sqrt{\mathbf{V} {(\Delta} U_k)}}\indist \mathcal{N} (0, 1), \quad  
    R_k:= \frac{\log(\Theta U_k)- \log(\Theta \tau)}{\sqrt{\mathbf{V} \left[\log{(\Theta U_k)}\right]}} \indist \mathcal{N} (0,1), \quad k\in\{1,\ldots,K\}.
\end{equation}
Variance of Net Benefit in stage $k$, $\mathbf{V} {(\Delta} U_k)$, can be easily derived based on formula (\ref{eq: VarDiffU}) by substituting $m_k, n_k$ for $m,n$, respectively. Also similarly, $\mathbf{V} {\left[\log{(\Theta U_k)}\right]}$ can be derived in the same manner.

\subsection{Joint Asymptotic Distribution of Sequential Net Benefit and Win Ratio}

In this section, we establish the joint asymptotic normality for the proposed sequential Net Benefit $\{D_k\}_{k=1}^K$ (NB) and sequential Win Ratio $\{R_k\}_{k=1}^K$ (WR), considering both the standardized versions (sequential SNBs and SWRs) and the corresponding non-standardized versions (sequential NSNBs and NSWRs). The following propositions demonstrate the properties of the derived Net Benefit and Win Ratio statistics in sequential design scenarios.

\begin{Proposition} \label{Prop: WinDiff}
\textit{Canonical Joint Distribution for Sequential Net Benefit}

Considering sequential standardized Net Benefit (seq-SNB) $\{D_k\}_{k=1}^K$ or their non-standardized versions (seq-NSNB), we have the following results:
\begin{enumerate} 
    \item \textit{(Independent Increment of Net Benefit)} For any pair of seq-SNBs $D_p, D_q$ at different stages $1 \le p < q \le K$, we have
    $$
    \mathbf{Cov}(D_p, D_q)=\sqrt{{\mathbf{V} (\Delta U_q)}/{\mathbf{V} (\Delta U_p)}}.
    $$
    For the non-standardized version, we have $\mathbf{Cov}(\Delta U_p, \Delta U_q) = \mathbf{V} (\Delta U_q).$

    \item \textit{(Joint Asymptotic Normality of Net Benefit)} Under Condition C1, for the seq-SNBs, we have 
    \begin{equation}
        \Big(D_1, D_2, \ldots, D_K\Big)^\top = \left(\frac{\Delta U_1 - \Delta \tau}{\sqrt{\mathbf{V} (\Delta U_1)}}, \frac{\Delta U_2 - \Delta \tau}{\sqrt{\mathbf{V} (\Delta U_2)}}, \cdots, \frac{\Delta U_K - \Delta \tau}{\sqrt{\mathbf{V} (\Delta U_K)}}\right)^\top \indist \mathcal{N}\left(\mathbf{0}, \mathbf{\Sigma}_{D1}\right),
    \end{equation}
    as $m_k - m_{k-1} \to \infty$ and $n_k - n_{k-1} \to \infty$ for $k = 1, 2, \cdots, K$. The asymptotic variance-covariance matrix $\mathbf{\Sigma}_{D1}$ is given by $\mathbf{\Sigma}_{D1} = (\sigma_{pq})_{K \times K},$ where the diagonal elements are 1, and the off-diagonal elements for $1 \le p < q \le K$ are defined as
    $$
    {\sigma}_{pq} = {\sigma}_{qp} = \sqrt{\frac{\left(\sum_{k=1}^q \eta_k\right)^{-1} \left(\xi^{10}_{ww} + \xi^{10}_{ll} - 2\xi^{10}_{wl}\right) + \left(\sum_{k=1}^q \gamma_k\right)^{-1} \left(\xi^{01}_{ww} + \xi^{01}_{ll} - 2\xi^{01}_{wl}\right)}{\left(\sum_{k=1}^p \eta_k\right)^{-1} \left(\xi^{10}_{ww} + \xi^{10}_{ll} - 2\xi^{10}_{wl}\right) + \left(\sum_{k=1}^p \gamma_k\right)^{-1} \left(\xi^{01}_{ww} + \xi^{01}_{ll} - 2\xi^{01}_{wl}\right)}}.
    $$
    For the seq-NSNB, we have 
    \begin{equation}
        \sqrt{N}\left(\Delta U_1 - \Delta \tau, \Delta U_2 - \Delta \tau, \ldots, \Delta U_K - \Delta \tau\right) \indist \mathcal{N}\left(\mathbf{0}, \mathbf{\Sigma}_{D2}\right),
    \end{equation}
    where the asymptotic variance-covariance matrix $\mathbf{\Sigma}_{D2}$ is given by $\mathbf{\Sigma}_{D2} = (\sigma_{pq}')_{K \times K},$ with elements for $1 \le p < q \le K$ defined as
    $
    \sigma_{pq}' = \sigma_{qp}' = \frac{\xi^{10}_{ww} + \xi^{10}_{ll} - 2\xi^{10}_{wl}}{\sum_{k=1}^q \eta_k} + \frac{\xi^{01}_{ww} + \xi^{01}_{ll} - 2\xi^{01}_{wl}}{\sum_{k=1}^q \gamma_k}.
    $
\end{enumerate}
$\hfill\square$
\end{Proposition}

\begin{Proposition} \label{Prop: WinRatio}
\textit{Canonical Joint Distribution for Sequential Win Ratio}

Considering  sequential standardized Win-Ratio statistics (seq-SWR) $R_1, R_2, \ldots, R_K$ or their non-standardized version (seq-NSWR), we have the following results: 
\begin{enumerate} 
    \item   \textit{(Asymptotic Independent Increment of Win Ratio)}  For any pair of seq-SWR $R_p, R_q$ at different stages $1 \le p < q \le K$, we have 
    $$\mathbf{Cov}(R_p, R_q)\approx\sqrt{{\mathbf{V} [\log(\Theta U_q)]}/{\mathbf{V} [\log(\Theta U_p)]}}.$$
    For seq-NSWR, we have $\mathbf{Cov}\left[\log (\Theta U_p), \log(\Theta U_q)\right] \approx \mathbf{V} \left[\log(\Theta U_q)\right].$
    
    \item \textit{(Joint Asymptotic Normality of Win Ratio)} Under Condition C1, For seq-SWR, we have
   \begin{equation}
   \begin{aligned}
        \left(R_1, R_2, \ldots, R_K\right)^\top &= \left(\frac{\log(\Theta U_1)- \log(\Theta \tau)}{\sqrt{\mathbf{V} \left[\log{(\Theta U_1)}\right]}}, \frac{\log(\Theta U_2)- \log(\Theta \tau)}{\sqrt{\mathbf{V} \left[\log{(\Theta U_2)}\right]}}, \ldots, \frac{\log(\Theta U_K)- \log(\Theta \tau)}{\sqrt{\mathbf{V} \left[\log{(\Theta U_K)}\right]}}\right)\\ &\indist \mathcal{N}\left(\mathbf{0},\mathbf{\Sigma}_{R1}\right),
   \end{aligned}
    \end{equation} 
    as $m_k-m_{k-1}\to\infty$ and $n_k-n_{k-1}\to\infty$ for $k=1,2,\cdots,K$. Asymptotic variance-covariance matrix $\mathbf{\Sigma}_{R1}$ is given by $\mathbf{\Sigma}_{R1} = (\sigma_{pq})_{K\times K},$ where the diagonal elements are 1, and off-diagonal elements for $1\le p<q\le K$, $\sigma_{pq}$ is defined as
    $$
    {\sigma}_{pq}={\sigma}_{qp}=\sqrt{{\Gamma^q}/{\Gamma^p}},
    $$
    where for simplicity, we define $\forall s, 1\leq s \leq K,$ 
  \begin{equation}
    \begin{aligned}
    \Gamma^s &= \frac{(\sum_{k=1}^s\eta_k)^{-1}\xi^{10}_{ww}+(\sum_{k=1}^s\gamma_k)^{-1}\xi^{01}_{ww}}{\tau^2_w} + \frac{(\sum_{k=1}^s\eta_k)^{-1}\xi^{10}_{ll}+(\sum_{k=1}^s\gamma_k)^{-1}\xi^{01}_{ll}}{\tau^2_l} \\
    &\quad -2 \frac{(\sum_{k=1}^s\eta_k)^{-1}\xi^{10}_{wl} + (\sum_{k=1}^s\gamma_k)^{-1}\xi^{01}_{wl}}{\tau_w\tau_l};
    \end{aligned}
\end{equation}

For seq-NSWR, we have 
\begin{equation}
    \sqrt{N}\Big( \log(\Theta U_1)-\log(\Theta\tau), \ldots, \log(\Theta U_K)-\log(\Theta\tau)\Big)\indist \mathcal{N}\left(\mathbf{0},\mathbf{\Sigma}_{R2}\right),
\end{equation}
where asymptotic variance-covariance matrix $\mathbf{\Sigma}_{R2}$ is given by $\mathbf{\Sigma}_{R2} = (\sigma_{pq}')_{K\times K},$ for $1\le p\leq q\le K$, we have
$\sigma_{pq}'=\sigma_{qp}'= \Gamma^{q}.$  
\end{enumerate}
\end{Proposition}
\begin{remark}
Both results for Part (1) of Proposition \ref{Prop: WinDiff} and \ref{Prop: WinRatio} are known as the independent increments of the sequential statistics by \cite{kim2020independent} and we have shown it also applies to our sequential Net Benefit and Win Ratio statistics. In addition,  results for Part (2) of Proposition \ref{Prop: WinDiff} and \ref{Prop: WinRatio} both help to establish the sequential test statistics in the design setting, such as Type-I error control and Power analysis. Detailed proof for Proposition \ref{Prop: WinDiff} and \ref{Prop: WinRatio} are provided in Appendix. $\hfill \square$
\end{remark}

After establishing the joint normality and independent increment properties, we consider the construction of hypothesis test statistics. The variance estimation method for both sequential SNBs and SWRs needs to be provided for our sequential statistics. Additionally, the asymptotic variance-covariance matrix must be estimated to ensure Type-I error control and power analysis.

For the sequential statistics, we need to estimate \(\xi^{10}_{uv}\), \(\xi^{01}_{uv}\), and \(\xi^{11}_{uv}\) for seq-SNBs and seq-NSNBs. Furthermore, we need to estimate \(\tau_w\) and \(\tau_l\) for SWRs and NSWRs. In our study, empirical estimators for \(\xi^{10}_{uv}\), \(\xi^{01}_{uv}\), \(\xi^{11}_{uv}\), \(\tau_w\), and \(\tau_l\) are provided in the Appendix. In this study, we test the hypothesis of the sequential study with the null hypothesis \( H_0: \tau_w = \tau_l \) against the one-sided alternative hypothesis \( H_\mathcal{A}: \tau_w > \tau_l \). Therefore, we propose  seq-SNBs and seq-SWRs test statistics corresponding to the \( K \) monitoring times under the null hypothesis as follows:

\begin{equation}
    \left\{\frac{\Delta U_k - 0}{\sqrt{\widetilde{\mathbf{V} (\Delta U_k)}}}\right\}_{k=1}^K \quad \text{and} \quad
    \left\{\frac{\log(\Theta U_k) - 0}{\sqrt{\widetilde{\mathbf{V} \left[\log(\Theta U_k)\right]}}}\right\}_{k=1}^K,
\end{equation}
where \(\widetilde{\mathbf{V} (\Delta U_k)}\) and \(\widetilde{\mathbf{V} \left[\log(\Theta U_k)\right]}\) are empirical estimates for \(\mathbf{V} (\Delta U_k)\) and \(\mathbf{V} \left[\log(\Theta U_k)\right]\), respectively, by plugin the empirical estimator of \(\xi^{10}_{uv}\), \(\xi^{01}_{uv}\), \(\xi^{11}_{uv}\), \(\tau_w\), and \(\tau_l\) to their corresponding estimation formula. In addition, empirical estimators for seriers of variance-covariance matrix $\mathbf{\Sigma}_{D1}$, $\mathbf{\Sigma}_{D2}$, $\mathbf{\Sigma}_{R1}$ and $\mathbf{\Sigma}_{R2}$ could also be derived. Non-standardized versions of sequential test statistics could also be developed. Details of empirical estimators are provided in the Web Supplementary  Material.

\section{Power Analysis and Sample Size Determination}\label{Sec: PowerANA}

\subsection{Type I Error Spending and Theoretical Power} \label{Sec: AlphaPower}
In this section, we present a framework for Type I error controlling and evaluating theoretical power, including a discussion on the alpha spending method and critical boundaries. The framework employed here followed \cite{lan:demets:1983}. As for the alpha spending function, \cite{kim:demets:1987} recommended several strictly convex and monotonically increasing alpha spending functions developed by \cite{lan:demets:1983} and \cite{kim:demets:1987}. In our study, we utilize  
$\alpha(t_k) = \alpha \times t_k^2$ for alpha spending function, where $t_k$ denotes the information fraction up to interim time point $k$, defined as $t_k = \frac{\mathbf{V}(\Delta U_K)}{\mathbf{V}(\Delta U_k)}$ for sequential NBs or $t_k = \frac{\mathbf{V} \left[\log(\Theta U_k)\right]}{\mathbf{V} \left[\log(\Theta U_k)\right]}$ for sequential WRs, for $k=1,2,\ldots, K$. This alpha spending function is derived from the Kim-Demets power family $\alpha(t,m) = \alpha\times t^m, \forall m>0$. As is shown in \cite{kim:demets:1987}, a strictly convex alpha spending function leads to a steady increase in partial Type I error, resulting in monotonically decreasing critical boundaries $\{c_k\}_{k=1}^K$. 

The partial Type I error values, or partial alpha values, at each monitoring time $1,2,\ldots, K$ are set as $\alpha_k = \alpha(t_k, 2) - \alpha (t_{k-1}, 2),$ with $\alpha (t_0, 2) = 0$ at the outset and $m = 2 $ for the family parameter. Critical boundaries can be estimated using the split alpha values by the alpha spending function under the null hypothesis $H_0$. Assuming a one-sided hypothesis test, we follow \cite{slud1982two} in constructing discrete sequential critical boundaries $\{c_k\}_{k=1}^K$. This is achieved by setting
\begin{equation}\label{Eq: CriticalBH0}
    \alpha _1 = \Pr \left(S_1^0 > c_{1}  \mid H_0 \right), \quad
    \alpha _k = \Pr \left(S_{1}^0 < c_{1}, \ldots, S_{k-1}^0 < c_{k-1} , S^0_{k} \ge c_k  \mid H_0 \right) ,
\end{equation}
for $k=2,3,\ldots,K$. Here, $\left(S^0_{1}, S^0_{2}, \ldots, S^0_K\right)^\top$ follows the K dimensional $\mathbf{MVN\left(0, \Sigma^0\right)}$ under the null hypothesis, with a zero mean vector $\mathbf{0_{K\times1}}$ and a variance-covariance matrix $\mathbf{\Sigma}^0$ could be valued by any one of previous four developed variance-covariance matrix, $\mathbf{\Sigma}_{D1}$, $\mathbf{\Sigma}_{D2}$, $\mathbf{\Sigma}_{R1}$ and $\mathbf{\Sigma}_{R2}$ under the $H_0$.

To evaluate the theoretical power, we employ the critical boundaries $\{c_k \}_{k=1}^K$ determined under the null hypothesis $H_0$ and compute the rejection probabilities $\{\mathrm{RP}_k\}_{k=1}^K$ across $K$ monitoring times. The rejection probabilities are defined as follows:
\begin{equation}\label{Eq: PowerH1}
    \begin{aligned}
\mathrm{RP}_1  = \Pr \left(S^A_{1} > c_{1} \mid H_\mathcal{A}\right), \quad  \mathrm{RP}_k  = \Pr \left(S^A_{1} < c_{1} , \ldots, S^A_{k-1} < c_{k-1} , S_{k} \ge c_k  \mid H_\mathcal{A}\right),
    \end{aligned}
\end{equation}
for $k=2,3, \ldots, K$. Here $\left(S^A_{1}, S^A_{2}, \ldots, S^A_K\right)^\top$ follows a K-dimensional multivariate Normal distribution $\boldsymbol{N}\left(\boldsymbol{\mu
}, \boldsymbol{\Sigma}^\mathcal{A}\right)$ under the alternative hypothesis $H_\mathcal{A}$ and $\boldsymbol{\mu
}$ is the estimand and it could be $\tau_w - \tau_l$ or $\log{(\tau_w/\tau_l)}$ under the $H_\mathcal{A}$. The sum of these $K$ rejection probabilities yields the overall theoretical power, i.e., 
$1-\beta = \sum_{k=1}^K \mathrm{RP}_k.
$

\subsection{Sample Size Determination}
For sample size determination, we need to specify the distribution of each outcome in $\mathcal{D}$ and $\mathcal{D}'$ further under $H_0$ and $H_\mathcal{A}$, respectively. Considering testing and study design, it is necessary to evaluate $\boldsymbol{\Sigma}^0$ (under $H_0$),  $\boldsymbol{\mu}$ (under $H_\mathcal{A}$) and $\boldsymbol{\Sigma}^\mathcal{A}$ (under $H_\mathcal{A}$) displayed in (\ref{Eq: CriticalBH0} and \ref{Eq: PowerH1}), both of which are critical for theoretical power calculation. First, $\boldsymbol{\mu}$ can be easily approximated by $\{U_{wk}, U_{lk}\}_{k=1}^K$ with very large data samples from the pre-specified distributions for $\mathcal{D}$ and $\mathcal{D}'$ under $H_\mathcal{A}$ by its empirical estimator $\widetilde{\boldsymbol{\mu}}$. Similarly, $\boldsymbol{\Sigma}^0$ and $\boldsymbol{\Sigma}^\mathcal{A}$ can be approximated by their empirical estimator $\widetilde{\boldsymbol{\Sigma}^0}$ and $\widetilde{\boldsymbol{\Sigma}^\mathcal{A}}$ (given Web Supplementary  Material) using very large generated $\mathcal{D}$ and $\mathcal{D}'$ data sample or called super-population as well under $H_0$ and $H_\mathcal{A}$, respectively.

It can be very tedious, if not impossible, to provide explicit formulas for a group sequential design as is possible for a fixed design. For simplicity, we assume that the total sample size is evenly split between the two arms and across $K$ stages. We propose an iterative approach to find the sample size $N$ for the proposed sequential design, as outlined in Algorithm~\ref{alg:1} when $K$ stages are planned. 

The basic idea behind this algorithm is to search for different sample sizes $N$ and identify the one that results in the required power. The key step in this process is determining the testing power for each sample size, which can be achieved through the aforementioned theoretical power evaluation approach. Furthermore, this algorithm can be extended to scenarios where the sample size is unevenly allocated between the two arms or across different interim stages.

\begin{algorithm}
\caption{Sample size determination algorithm for K-stage design }\label{alg:1}
\begin{algorithmic}
\Require \\
\begin{itemize}
\item  Total alpha level $\alpha$
\item  Alpha spending method
\item  Theoretical power level $p_0$
\item  Marginal Distribution of each outcome and copula method for the joint distribution  
\end{itemize}
  \vspace{5pt}
\State Estimate $\widetilde{\tau_w}, \widetilde{\tau_l},$ and $\widetilde{\Delta \tau}$ or $\widetilde{\Theta \tau}$ $\gets$ Super-populations under $ H_0, H_\mathcal{A}$ 

\State Estimate $ \{\widetilde{\xi^{uv}_{01}}, \widetilde{\xi^{uv}_{10}}, \widetilde{\xi^{uv}_{11}}\}, u,v \in \{w, l\} \text{, and }  \widetilde{\boldsymbol{\Sigma}^0}, \widetilde{\boldsymbol{\Sigma}^\mathcal{A}} \gets$ Super-populations under $ H_0, H_\mathcal{A}$

\vspace{5pt}
\State {\underline{Initial total size}}:~~ $N \gets n_0$ (a small multiple of $2K$)
 \vspace{5pt}
\State {\underline{Initial power}}:~~ $p \gets 0$
 \vspace{5pt}
\While {$p<p_0$}
\vspace{5pt}
    \State $N \gets N+2K$
    \State $\{\widetilde{\mathbf{V}(\Delta U_k)}\}_{k=1}^K  \text{ or }\{\widetilde{\mathbf{V}[\log(\Theta U_k)]}\}_{k=1}^K \gets  $ Under $ H_0, H_\mathcal{A} \text{ and } N$
    \State $\{\alpha_k\}_{k=1}^K \gets t_{k,N}=\frac{\widetilde{\mathbf{V}(\Delta U_K)}}{\widetilde{\mathbf{V}(\Delta U_k)}}$ or $\frac{\widetilde{\mathbf{V}[\log(\Theta U_K)]}}{\widetilde{\mathbf{V}[\log(\Theta U_k)]}}$ under $H_0$
    \vspace{5pt}
    \State {\underline{Critical boundaries}}:~~$\{c_k\}_{k=1}^K \gets \{\alpha_k\}_{k=1}^K \text{~and~} \widetilde{\boldsymbol{\Sigma}^0}$ (based on joint normality)
    \vspace{5pt}
    \State $\boldsymbol{\mu} \gets \left(\frac{\widetilde{\Delta\tau}}{\widetilde{\mathbf{V}(\Delta U_1)}}, \ldots, \frac{\widetilde{\Delta \tau}}{\widetilde{\mathbf{V}(\Delta U_K)}}\right)^\top$ or $\left(\frac{\log \widetilde{(\Theta \tau)}}{\widetilde{\mathbf{V}[\log (\Theta U_1)]}}, \ldots, \frac{\log \widetilde{(\Theta \tau)}}{\widetilde{\mathbf{V}[\log (\Theta U_K)]}}\right)^\top$ under $H_\mathcal{A}$
    \vspace{5pt}
    \State {\underline{Theoretical Power at $N$}}:~~$p \gets \{c_k\}_{k=1}^K, \boldsymbol{\mu} \text{~and~} \widetilde{\boldsymbol{\Sigma}^\mathcal{A}}$ (based on joint normality)
    \vspace{5pt}
\EndWhile
\end{algorithmic}
\end{algorithm}

\section{Numerical Analysis}\label{Sec: NumericalAna}

In this section, we conduct a numerical analysis to mimic the scenario of a cardiovascular trial, whose primary outcome typically includes an HCE of death $D_1$, hospitalization for heart failure $D_2$, and New York Heart Association (NYHA) Functional Classification $D_3$ (\citeauthor{james2024dapagliflozin}, \citeyear{james2024dapagliflozin}). The first two outcomes are time-to-event, and the last is four levels of ordinal outcome, with small to large representing deterioration. Following the notation above, we denote the HCE in the treatment arm as $\mathcal{D} = \{D_1, D_2, D_3\}=\{(Y_1, \delta_1), (Y_2, \delta_2), D_3\}.$ We compare the fixed method and sequential method in these scenario settings, according to but not limited to power level and Type-I error. For simplicity, we assume an overall number of stages $K=3$ and an allocation ratio for each stage and arm we have $\eta_k=\gamma_k=1/6, k=1,2,3.$ 

Considering the correlation between outcomes in the HCE, a trivariate Clayton copula is applied to generate the outcomes in HCE, where the copula parameter, denoted as \(\rho\), is derived from a common Kendall's \(\tau\) across all pairwise comparisons \citep{oakes1989bivariate, hofert2018elements}. Specifically, the copula parameter is calculated using the relationship \(\rho = \frac{2 \tau}{1 - \tau}\), which facilitates the modeling of right tail dependence. This assumption implies that the strength of dependence between endpoints, as measured by Kendall's \(\tau\), is uniform across all three dimensions of the copula. Suppose survival outcome death $D_1$ and hospitalization $D_2$ both follow exponential distributions with parameters $\lambda_{1Z}$ and $\lambda_{2Z}$. Let $\lambda_{qZ}=\lambda_{q}\exp\{-\beta_qZ\}$ be the hazard rate for the $q-$th event, where $Z=1$ if subject in the treatment group and $Z=0$ on the standard group. For ordinal outcome ${D}_3$, we consider following proportional odds model
\begin{equation}
\label{eq: PropOddOrdinal}
    \begin{aligned}
        \log \frac{\sum_{j=l}^4 p_j^{(1)}}{\sum_{j=1}^{l-1} p_j^{(1)}}=  \log \frac{\sum_{j=l}^4 p_j^{(0)}}{\sum_{j=1}^{l-1} p_j^{(0)}} -\beta_3,
    \end{aligned}
\end{equation}
for $l=2,3,4$, where $p_j^{(Z)}$ denotes as population proportion of level $l$ in group $Z$. Independent of time-to-event outcomes, we assume that the censoring variable $C$  follows a uniform distribution on the interval $[0, c]$. To be noted, in our study, $C$ could be extended to follow any kind of distribution, like exponential distribution or uniform distribution $[c_1, c_2]$. 

The hypotheses are established as $H_0: \beta_1 = \beta_2 = \beta_3 = 0$ and $H_\mathcal{A}: \text{At least one } \beta_q > 0 \text{ for } q \in \{1, 2, 3\}$. Throughout the simulation, the parameters $\lambda_1 = 0.08, \lambda_2 = 0.1 , \{p_j^{(0)}\}_{j=1}^4 = 0.25, \beta_3 = 0.25,$ and $\tau= 0.3$ were held constant, while $\beta_1$, $\beta_2$, and $c$ were varied in different simulation settings. We test the scenarios combination where $\beta_1, \beta_2 = 0.2$ or 0.4 under the $H_\mathcal{A}$ and $c = 12$ or 24. Type I error was controlled at the 0.05 level, with the power fixed at 90\%. Four methods were compared based on their theoretical and attained power or Type I error levels: seq-SNB, seq-SWR, fix-SNB, and fix-SWR. The last two represent fixed designs with standardized NB and WR. For simplicity, the sample size across different methods was kept the same in the same scenario and was determined by seq-SNB using Algorithm \ref{alg:1}. For each scenario, 10,000 Monto Carlo simulations are generated to obtain the attained corresponding Type-I error and power level. For comparison, the bias between attained and theoretical Type-I or power level was also reported.

Simulation results are shown in Table \ref{Tab: TypeIError} (Type I error control) and Table \ref{Tab: PowerANA} (Power analysis at the 90\% level). Firstly, bias for both Type I error rates and power levels is minimal across all methods and parameter scenarios, indicating that the asymptotic theory is well-supported. In Table \ref{Tab: PowerANA}, fixed-design methods exhibit slightly higher power than sequential designs. As a result, they require a smaller maximum sample size (MSS), though not necessarily a smaller average sample size (ASN). When accounting for early stopping, sequential methods demonstrate a markedly reduced ASN. Regarding the different parameter settings under $H_\mathcal{A}$, the larger the value of $\beta_q$, the smaller the sample size needed to achieve the corresponding power level. Finally, there is no clear difference regarding the attained power level between seq-SNBs and seq-SWRs. 

\begin{table}
\centering
\begin{tabular}{ccc||cc||c|c}
\hline\hline\hline
$\beta_1$ & $\beta_2$ & c & Method & MSS & Attained Type I Rate (\%) & Type I Error Bias (\%) \\ \hline\hline\hline
\multirow{16}{*}{0} & \multirow{8}{*}{0} & \multirow{4}{*}{12} & seq-SNBs & \multirow{4}{*}{432} & (0.60, 2.36, 5.02) & (0.04, 0.14, 0.02) \\
 &  &  & seq-SWRs &  & (0.57, 2.22, 5.03) & (0.01, 0.00, 0.03) \\
 &  &  & fix-SNB &  & 5.20 & 0.20 \\
 &  &  & fix-SWR &  & 5.17 & 0.17 \\ \cline{3-7}
 &  & \multirow{4}{*}{24} & seq-SNBs & \multirow{4}{*}{456} & (0.57, 2.38, 5.06) & (0.01, 0.16, 0.06) \\
 &  &  & seq-SWRs &  & (0.59, 2.31, 5.07) & (0.03, 0.09, 0.07) \\
 &  &  & fix-SNB &  & 5.00 & 0.00 \\
 &  &  & fix-SWR &  & 5.07 & 0.07 \\ \cline{2-7}
 & \multirow{8}{*}{0} & \multirow{4}{*}{12} & seq-SNBs & \multirow{4}{*}{279} & (0.79, 2.36, 4.99) & (0.23, 0.14, -0.01) \\
 &  &  & seq-SWRs &  & (0.53, 2.30, 5.04) & (-0.03, 0.08, 0.04) \\
 &  &  & fix-SNB &  & 5.24 & 0.24 \\
 &  &  & fix-SWR &  & 5.11 & 0.11 \\ \cline{3-7}
 &  & \multirow{4}{*}{24} & seq-SNBs & \multirow{4}{*}{303} & (0.64, 2.26, 5.06) & (0.08, 0.04, 0.06) \\
 &  &  & seq-SWRs &  & (0.57, 2.25, 5.37) & (0.01, 0.03, 0.37) \\
 &  &  & fix-SNB &  & 4.74 & -0.26 \\
 &  &  & fix-SWR &  & 5.43 & 0.43 \\ \hline\hline
\multirow{16}{*}{0} & \multirow{8}{*}{0} & \multirow{4}{*}{12} & seq-SNBs & \multirow{4}{*}{228} & (0.66, 2.69, 5.53) & (0.10, 0.47, 0.53) \\
 &  &  & seq-SWRs &  & (0.51, 2.19, 4.99) & (-0.05, -0.03, -0.01) \\
 &  &  & fix-SNB &  & 4.78 & -0.22 \\
 &  &  & fix-SWR &  & 5.55 & 0.55 \\ \cline{3-7}
 &  & \multirow{4}{*}{24} & seq-SNBs & \multirow{4}{*}{189} & (0.94, 2.59, 5.64) & (0.38, 0.37, 0.64) \\
 &  &  & seq-SWRs &  & (0.54, 2.19, 5.04) & (-0.02, -0.03, 0.04) \\
 &  &  & fix-SNB &  & 5.21 & 0.21 \\
 &  &  & fix-SWR &  & 4.73 & -0.27 \\ \cline{2-7}
 & \multirow{8}{*}{0} & \multirow{4}{*}{12} & seq-SNBs & \multirow{4}{*}{165} & (0.74, 2.63, 5.82) & (0.18, 0.41, 0.82) \\
 &  &  & seq-SWRs &  & (0.58, 2.30, 5.15) & (0.02, 0.08, 0.15) \\
 &  &  & fix-SNB &  & 5.31 & 0.31 \\
 &  &  & fix-SWR &  & 5.07 & 0.07 \\ \cline{3-7}
 &  & \multirow{4}{*}{24} & seq-SNBs & \multirow{4}{*}{141} & (0.90, 2.63, 5.33) & (0.34, 0.41, 0.33) \\
 &  &  & seq-SWRs &  & (0.71, 2.34, 5.16) & (0.15, 0.12, 0.16) \\
 &  &  & fix-SNB &  & 5.60 & 0.60 \\
 &  &  & fix-SWR &  & 5.02 & 0.02 \\ \hline\hline\hline
\end{tabular}

~

\caption{Type I level control in each interim analysis under $H_0$ with $\alpha = 0.05$ level. MSS refers to the maximum sample size in a single arm. The "Type I Error Bias" is defined as the difference between the attained and theoretical Type I error rates at each stage. For the sequential methods (seq-SNBs and seq-SWRs), the values in parentheses under "Attained Type I Rate" and "Type I Error Bias" represent the results at Interim 1, Interim 2, and Final analyses, respectively. Specifically, each set of three numbers corresponds to the attained Type I error rate or bias at each of these three analysis points. }
\label{Tab: TypeIError}
\end{table}

\begin{table}
\centering
\begin{tabular}{ccc||cc||c|c}
\hline \hline \hline
$\beta_1$ & $\beta_2$ & c & Method & MSS (ASN) & Attained Power Level (\%) & Bias of Power (\%) \\ \hline\hline
\multirow{16}{*}{0.2} & \multirow{8}{*}{0.2} & \multirow{4}{*}{12} & seq-SNBs & 432 (352) & (22.49, 66.61, 90.91) & (1.57, 0.94, 0.60) \\
                      &                      &                     & seq-SWRs & 432 (352) & (21.15, 66.54, 90.92) & (0.60, 1.55, 0.96) \\
                      &                      &                     & fix-SNB & 432       & 91.44 & 0.17 \\
                      &                      &                     & fix-SWR & 432       & 91.49 & 0.59 \\ \cline{3-7}
                      &                      & \multirow{4}{*}{24} & seq-SNBs & 456 (372) & (22.22, 66.37, 90.98) & (1.43, 0.94, 0.86) \\
                      &                      &                     & seq-SWRs & 456 (371) & (20.48, 65.99, 90.77) & (0.05, 1.23, 0.98) \\
                      &                      &                     & fix-SNB & 456       & 91.70 & 0.56 \\
                      &                      &                     & fix-SWR & 456       & 91.42 & 0.65 \\ \cline{2-7}
                      & \multirow{8}{*}{0.4} & \multirow{4}{*}{12} & seq-SNBs & 279 (228) & (22.59, 65.42, 90.01) & (1.77, -0.12, -0.25) \\
                      &                      &                     & seq-SWRs & 279 (227) & (21.11, 65.84, 90.92) & (0.87, 1.36, 1.29) \\
                      &                      &                     & fix-SNB & 279       & 91.44 & 0.23 \\
                      &                      &                     & fix-SWR & 279       & 91.39 & 0.76 \\ \cline{3-7}
                      &                      & \multirow{4}{*}{24} & seq-SNBs & 303 (247) & (21.39, 64.81, 89.35) & (0.64, -0.59, -0.81) \\
                      &                      &                     & seq-SWRs & 303 (247) & (19.89, 64.88, 89.35) & (-0.31, 0.50, -0.23) \\
                      &                      &                     & fix-SNB & 303       & 90.70 & -0.43 \\
                      &                      &                     & fix-SWR & 303       & 90.66 & 0.08 \\ \hline \hline
\multirow{16}{*}{0.4} & \multirow{8}{*}{0.2} & \multirow{4}{*}{12} & seq-SNBs & 228 (186) & (22.47, 65.44, 89.95) & (1.68, -0.09, -0.30) \\
                      &                      &                     & seq-SWRs & 228 (186) & (20.28, 64.46, 89.37) & (0.16, 0.16, -0.15) \\
                      &                      &                     & fix-SNB & 228       & 90.55 & -0.66 \\
                      &                      &                     & fix-SWR & 228       & 90.04 & -0.50 \\ \cline{3-7}
                      &                      & \multirow{4}{*}{24} & seq-SNBs & 189 (154) & (22.84, 66.20, 90.64) & (2.33, 1.14, 0.63) \\
                      &                      &                     & seq-SWRs & 189 (154) & (19.95, 63.78, 89.93) & (0.25, 0.24, 0.83) \\
                      &                      &                     & fix-SNB & 189       & 90.98 & 0.01 \\
                      &                      &                     & fix-SWR & 189       & 91.16 & 1.03 \\ \cline{2-7}
                      & \multirow{8}{*}{0.4} & 12                  & seq-SNBs & 165 (135) & (23.85, 66.76, 90.80) & (3.00, 1.05, 0.42) \\
                      &                      &                     & seq-SWRs & 165 (134) & (20.70, 66.20, 90.20) & (0.79, 2.22, 0.83) \\
                      &                      &                     & fix-SNB & 165       & 90.83 & -0.50 \\
                      &                      &                     & fix-SWR & 165       & 91.16 & 0.78 \\ \cline{3-7}
                      &                      & \multirow{4}{*}{24} & seq-SNBs & 141 (115) & (23.19, 65.21, 89.72) & (2.67, 0.04, -0.36) \\
                      &                      &                     & seq-SWRs & 141 (115) & (19.35, 63.06, 89.26) & (-0.07, -0.04, 0.41) \\
                      &                      &                     & fix-SNB & 141       & 90.28 & -0.77 \\
                      &                      &                     & fix-SWR & 141       & 91.08 & 1.19 \\ \hline \hline \hline
\end{tabular}

~

\caption{Power Analysis at 90\% level between Fixed and Sequential Methods. MSS refers to the maximum sample size in a single arm; ASN in the parentheses refers to the average sample size. The "Bias of Power" is defined as the difference between the attained and theoretical power levels at each stage. For the sequential methods (seq-SNBs and seq-SWRs), the values in parentheses under "Attained Power Level" and "Bias of Power" represent the results at Interim 1, Interim 2, and Final analyses, respectively. Specifically, each set of three numbers corresponds to the attained power or bias at each of these three analysis points. } \label{Tab: PowerANA}
\end{table}

\section{Real Data Analysis}\label{Sec: RealANA}
In this section, we apply our developed design based on sequentially derived win statistics to a real-world example from the HF-ACTION, a cardiovascular study conducted by \cite{o2009efficacy}. The primary objective was to evaluate the effect of adding exercise training to the usual patient care on the composite endpoint of all-cause hospitalization $D_1$  and death $D_2$. In our study, we focus on a high-risk subgroup consisting of 426 non-ischemic patients with baseline cardio-pulmonary exercise tests no longer than nine minutes before reporting discomfort. In this subgroup, 205 patients were assigned to receive exercise training in addition to usual care and 221 were assigned to receive usual care alone. Data can be found in R package \textit{rmt}. In our study,  we utilize an HCE, denoted as $\mathcal{D}=\{D_1, D_2\}=\{(Y_1, \delta_1), (Y_2, \delta_2)\}$, where $Y_q$ represents observed event time and $\delta_q$ is observed indicator for endpoint $q$, prioritized by clinical priority. The first hospitalization event was used in the HCE. 

\begin{figure}
    \caption{Hierarchical Composite Outcome assessed by the Win Ratio and Net Benefit Method. The left panel reports the Win and Loss proportion for Death + Hospitalization (HCE) vs Death Only. The middle panel reports the point estimate of NB as well as its 95\% CI. The right Panel reports the point estimate of WR as well as its 95\% CI. }
    \label{fig:HFACTION}
    \centering\includegraphics[width=1\linewidth]{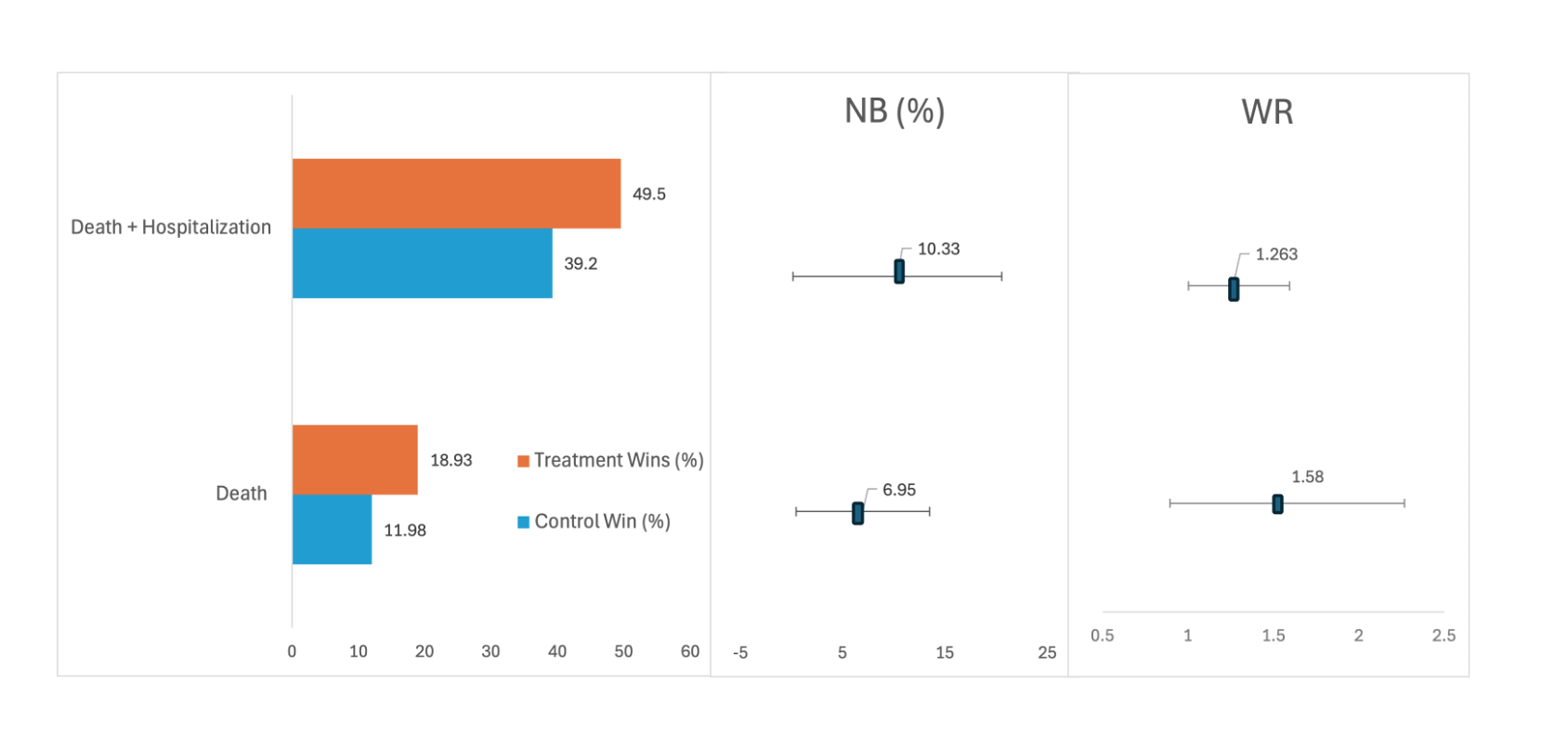}
\end{figure}

Analysis based on fixed design method is reported in Figure \ref{fig:HFACTION} and details can be found in Table \ref{tab: HCE_Real}. The HCE including events of death and hospitalization resulted in 49.50\% wins for treatment and 39.20\% for control (Net Benefit, 10.33\%; 95\% confidence interval [CI], 0.12\% to 20.53\%. Win Ratio, 1.265; 95\% CI, 1.001 to 1.594). The endpoint with death only resulted in 18.93\% wins for treatment and 11.98\% for control (Net Benefit, 6.95\%; 95\% [CI], 4.14\% to 13.48\%. Win Ratio, 1.580; 95\% CI, 0.895 to 2.265). 

\begin{table}
    \centering
    \begin{tabular}{c|c|c|c|c|c}\hline\hline\hline
        HCE Components & Wins& Losses& Ties& NB (95\% CI) & WR (95\% CI)\\
        \hline\hline\hline
         Death & 18.93\% & 11.98\%& 69.09\% & 6.95\% (4.14\%, 13.48\%) & 1.580 (0.895, 2.265)\\
         Death + Hospitalization & 49.50\%& 39.20\%& 11.26\%& 10.33\% (0.12\%, 20.53\%) & 1.263 (1.001, 1.594)\\
         \hline\hline\hline
    \end{tabular}
    \caption{Hierarchical Composite Outcomes assessed by the Win Ratio and Net Benefit Method for HF-ACTION data set.}
    \label{tab: HCE_Real}
\end{table}

Next, we compare seq-SNBs, seq-SWRs, fix-SNB, and fix-SWR based on the HF-ACTION data for theoretical power and average sample size. Empirical estimators for $\xi^{10}_{uv}$, $\xi^{01}_{uv}$, $\xi^{11}_{uv}$, $\tau_w$, and $\tau_l$ were derived from the HF-ACTION data. The asymptotic variance $\mathbf{\Sigma}$ for each method was estimated via plug-in empirical estimators. Critical boundaries, theoretical power, and average sample size based on the estimated variance for each method are reported in Table~\ref{tab: HFACTION}. For each method, theoretical power is calculated based on the actual sample size from this subgroup. Accordingly, the average sample size is decided based on the maximum sample size and theoretical power for each stage. From the results in Table~\ref{tab: HFACTION}, both sequential methods have a relatively smaller average sample size than both fixed design methods while maintaining the power level.

\begin{table}
    \centering

    \begin{tabular}{c|c|c|c}
    \hline \hline \hline
         Method &  ASN & Critical Boundaries & Theoretical Power Level (\%)\\
         \hline \hline \hline
        seq-SNBs & 367 & (2.54, 2.07, 1.74) & (7.94, 33.44, 61.49)\\
        seq-SWRs&  368& (2.54, 2.07, 1.73) & (7.92, 33.01, 61.01)\\
         fix-SNB &  426& 1.64 & 63.27\\
         fix-SWR &  426&  1.64 & 62.71\\
    \hline \hline \hline
    \end{tabular}
    
    ~
    
       \caption{Power analysis for HF-ACTION study with $\alpha = 0.05$ with maximum sample size 426. MSS refers to the maximum sample size in the control arm and treatment arm; ASN refers to the average maximum sample size in the control and treatment arm. In fixed design, ASN and MSS are the same}
    \label{tab: HFACTION}
\end{table}



\section{Concluding Remarks} \label{Sec: Conclusion}

The Win Ratio has been gaining traction as an analytical method in biomedical research due to its ability to incorporate study outcomes of different types with a clinical priority order into a single composite endpoint. This approach eliminates the need for multiple testing adjustments and provides a unified interpretation of treatment effects, regardless of the outcome data type. In this study, we developed a sequential design framework based on the Win Ratio and Net Benefits test statistics. Our proposed method effectively controls the Type I error rate while achieving the desired power. Furthermore, by allowing for early determination of treatment efficacy or futility, the sequential design requires a smaller average sample size compared to the fixed design, as expected, for both Win Ratio and Net Benefit test statistics.

One limitation of this study pertains to the estimand for the Win Ratio, which can vary depending on the censoring or follow-up distribution. A more natural estimand in such cases would be the Win Ratio, assuming all patients were followed for the same duration \citep{mao2024defining, oakes2016win}. To address the issue of varying censoring distributions, our sequential study censors participants recruited in each stage after each interim analysis, meaning no additional follow-up data is collected for these individuals. As a result, pairwise comparisons for constructing the Win-statistics are limited to subjects with the same follow-up distribution across different groups. 

A more practical alternative would be to avoid administrative censoring after interim analyses and to continue following up to the next stage. This approach would allow the collection of long-term information for each participant. However, one must carefully consider whether comparing patients with differing follow-up times is meaningful. Moreover, the pairwise comparison between the same pair of subjects can vary across different interim analyses due to differences in follow-up times. As a result, the joint canonical distribution may not hold for the sequential Win Ratio or Net Benefit, potentially undermining the applicability of a sequential design in this context.

\section{Data availability statement}
The real study data that support the findings in this paper are openly available on the Comprehensive R Archive Network (CRAN; https://cran.r-project.org/web/packages/rmt/index.html).

\bibliographystyle{biorefs}
\bibliography{refs}

\newpage

\section*{Appendix}
\section{Variance for Derived Win-statistics}

\subsection{Variance of Win Ratio Statistics} \label{App: VRatioU}
In this subsection, we derive the asymptotic variance of the Win Ratio. The below derivation process follows a similar idea from \cite{bergemann4450923group}. First, we review the results of Taylor expansions for the moments of functions of random variables. Given $\mu_X$ and $\sigma^2_X$ as the mean and the variance of a random variable $X$, respectively. According to \cite{wolter2007introduction}, the Taylor expansion of the expected value or first moment of $f(X)$ can be found as 
\[
\mathbf{E}[f(X)] \approx f(\mu_X) + f'(\mu_X)\mathbf{E}[X-\mu_X] + \frac{1}{2}f''(\mu_X)\mathbf{E}[(X-\mu_X)^2]=f(\mu_X)+\frac{f''(\mu_X)}{2}\sigma_X^2
\]
Suppose $f(x)=\exp{(x)}$ and the Taylor expansion of first moment of $\exp{(x)}$ yields 
\[\mathbf{E}[f(X)] \approx \exp{(\mu_X)} + \frac{1}{2}\exp{(\mu_X)}\sigma_X^2\approx \exp{\left\{\mu_X+\frac{1}{2}\sigma_X^2\right\}}. \]
Mentioned that there are tiny mistakes in \cite{bergemann4450923group} derivation of this part and we corrected them in our results. Then, we need to apply the results of Taylor's expansion again to simplify the above results. Considering arbitrary two random variables $X_1,X_2$, denote $Y_i=\log{(X_i)}$ for random variable transformation of $X_i, i=1,2$. 
\begin{align*}
    \frac{\mathbf{Cov}(X_1,X_2)}{\mathbf{E}X_1\mathbf{E}X_2}+1 &= \frac{\mathbf{E}(X_1X_2)}{\mathbf{E}X_1\mathbf{E}X_2} = \frac{\mathbf{E}[\exp{(Y_1+Y_2})]}{\mathbf{E}[\exp{(Y_1)}]\mathbf{E}[\exp{(Y_2)}]}\\
    &\approx \frac{\exp{\left\{\mathbf{E}(Y_1) + \mathbf{E}(Y_2) + \frac{1}{2}\mathbf{V} (Y_1+Y_2)\right\}}}
    {\exp{\left\{\mathbf{E}(Y_1) + \frac{1}{2}\mathbf{V} (Y_1)\right\}} \exp{\left\{\mathbf{E}(Y_2) + \frac{1}{2}\mathbf{V} (Y_2)\right\}}} \tag{By Taylor expansion }\\
    &=\exp{\{\mathbf{Cov}(Y_1,Y_2)\}} = \exp{\left\{\mathbf{Cov}\left(\log{X_1},\log{X_2}\right)\right\}}.
\end{align*}
Thus, we have the below approximation
$$
{\mathbf{Cov}\left(\log{X_1},\log{X_2}\right)}\approx \log\left(\frac{\mathbf{Cov}(X_1,X_2)}{\mathbf{E}X_1\mathbf{E}X_2}+1\right)\approx \frac{\mathbf{Cov}(X_1,X_2)}{\mathbf{E}X_1\mathbf{E}X_2}.
$$

Using above results, we can derive the approximation of $\mathbf{V}[\log(\Theta U)] = \mathbf{V}[\log U_w] + \mathbf{V}[\log U_l] - 2 \mathbf{Cov}[\log (U_w, U_l)].$
For $u=w,l$, we have 
\begin{align*}
    \mathbf{V}\left[\log U_u\right]\approx\frac{\mathbf{V}(U_u)}{\tau_u^2} = \frac{\frac{n-1}{mn}\xi_{10}^{uu} + \frac{m-1}{mn}\xi_{01}^{uu}+\frac{1}{mn}\xi_{11}^{uu}}{\tau_u^2},
\end{align*}
and also
\begin{align*}
    \mathbf{Cov}\left[\log (U_w), \log(U_l)\right]\approx \frac{\mathbf{Cov}(U_w, U_l)}{\tau_w\tau_l} =\frac{\frac{n-1}{mn}\xi^{10}_{wl} + \frac{m-1}{mn}\xi^{01}_{wl} + \frac{1}{mn}\xi_{11}^{wl}}{\tau_w\tau_l}.
\end{align*}

\subsection{Empirical Estimators of Variance}
In this subsection, we present the consistent empirical estimators for \(\xi^{10}_{uv}\), \(\xi^{01}_{uv}\), \(\xi^{11}_{uv}\), \(\tau_w\), and \(\tau_l\), which are key components of the variance for both the Win Ratio and Net Benefit. Firstly, the empirical estimators of \(\tau_w\) and \(\tau_l\) are given by their respective sample means, denoted as \(\widetilde{\tau_w}\) and \(\widetilde{\tau_l}\). For \(u \in \{w, l\}\), the estimator is defined as
\[
\widetilde{\tau_u} = \frac{1}{mn}\sum_{i=1}^m\sum_{j=1}^n\varphi_u\left(\mathcal{D}_i; \mathcal{D}'_j\right),
\]
which serves as a consistent estimator for \(\tau_u\).

Next, we derive the empirical estimators for covariance estimators. Firstly, we take $\xi^{10}_{uv}$ as an example to illustrate its empirical estimator. 
$$\xi^{10}_{uv}= \mathbf{Cov}\left[\varphi_u\left(\mathcal{D}_1; \mathcal{D}'_1\right), \varphi_v\left(\mathcal{D}_1; \mathcal{D}'_2\right)\right] = \mathbf{E}\left[\varphi_u\left(\mathcal{D}_1; \mathcal{D}'_1\right) \varphi_v\left(\mathcal{D}_1; \mathcal{D}'_2\right)\right] -\tau_u\times\tau_v .
$$
We have already proposed the consistent estimator for $\tau_w,$ and $\tau_l,$ respectively. Thus, the consistent estimator for $\tau_w \times\tau_l$ is $\widetilde{\tau_w}\times\widetilde{\tau_l}.$ Denote
\[
\varphi_{uv}^{10}\left(\mathcal{D}_i; \mathcal{D}'_{j_1}, \mathcal{D}'_{j_2}\right) = \frac{1}{2}\times
    \left[\varphi_u\left(\mathcal{D}_i; \mathcal{D}'_{j_1}\right)\varphi_v\left(\mathcal{D}_i; \mathcal{D}'_{j_2}\right) + \varphi_u\left(\mathcal{D}_i; \mathcal{D}'_{j_2}\right)\varphi_v\left(\mathcal{D}_i; \mathcal{D}'_{j_1}\right) 
\right],
\]
and the U-statistics with kernel $\varphi^{10}_{uv}\left(\mathcal{D}_i; \mathcal{D}'_{j_1}, \mathcal{D}'_{j_2}\right)$ is a consistent estimator for $\mathbf{E}\left[\varphi_u\left(\mathcal{D}_1; \mathcal{D}'_1\right) \varphi_v\left(\mathcal{D}_1; \mathcal{D}'_2\right)\right].$ Thus, we proposed the consistent estimator for $\xi^{10}_{uv}$ as 
\[
\widetilde{\xi^{10}_{uv}} = \frac{1}{mn(n-1)}\sum_{i=1}^m\sum_{j=1}^n\sum_{j'=1, ~j'\ne j}^n \varphi^{10}_{uv}(\mathcal{D}_i; \mathcal{D}'_j, \mathcal{D}'_{j'}) - \widetilde{\tau_u}\times\widetilde{\tau_v}.
\]
Similarly, consistent empirical covariance estimators for  \(\xi^{01}_{uv}\), and  \(\xi^{11}_{uv}\) are
\[
\widetilde{\xi^{01}_{uv}} = \frac{1}{m(m-1)n} \sum_{i_1=1}^m \sum_{i_2=1, i'\ne i}^m\sum_{j'=1}^n \varphi^{01}_{uv}(\mathcal{D}_i, \mathcal{D}'_i; \mathcal{D}'_{j}) - \widetilde{\tau_u}\times\widetilde{\tau_v},
\]
and
\[
\widetilde{\xi^{11}_{uv}}= \frac{1}{mn}\sum_{i=1}^m\sum_{j=1}^n \varphi^{11}_{uv}(\mathcal{D}_i; \mathcal{D}'_{j}) - \widetilde{\tau_u}\times\widetilde{\tau_v},
\]
where
\[
\varphi^{01}_{uv}(\mathcal{D}_i, \mathcal{D}_{i'};  \mathcal{D}'_{j}) =  \frac{1}{2}\left[\varphi_u\left(\mathcal{D}_i; \mathcal{D}'_j\right)\varphi_v\left(\mathcal{D}_{i'}; \mathcal{D}'_{j}\right) + \varphi_u\left(\mathcal{D}_{i'}; \mathcal{D}'_j\right)\varphi_v\left(\mathcal{D}_i; \mathcal{D}'_j\right)\right],
\]
and 
\[
\varphi^{11}_{uv}(\mathcal{D}_i; \mathcal{D}'_{j}) = \varphi_u\left(\mathcal{D}_i; \mathcal{D}'_j\right)\varphi_v\left(\mathcal{D}_{i}; \mathcal{D}'_{j}\right).
\]
Here we plug in $\tau_w$ and $\tau_l$ with $\widetilde{\tau_w}$ and $\widetilde{\tau_l}$ as their corresponding empirical estimators. With these empirical estimators, empirical estimators for $\sqrt{\mathbf{V} {(\Delta} U)}$, $\mathbf{V} \left[\log{(\Theta U)}\right]$ and asymptotic variance-covariance as shown in Proposition 1 and 2 could be derived.

\section{Proof of Lemma 1}\label{App: Lemma1}

The projection of  $U_{\nu k} -\tau_\nu$ onto Hilbert Space $\mathcal{H}$ can be written as $\hat{U}_{\nu k}$, where 
\begin{equation}
    \begin{aligned}
        \hat{U}_{\nu k} &= \sum_{p=1}^{n_k}\mathbf{E}\left[ U_{\nu k}-\tau_\nu \mid\mathcal{D}_p \right]  + \sum_{q=1}^{m_k} \mathbf{E}\left[ U_{\nu k} - \tau_\nu \mid\mathcal{D}'_q \right] \\
        & = \sum_{p=1}^{m_k} \mathbf{E}\left[ \frac{1}{m_kn_k}\sum_{i=1}^{m_k}\sum_{j=1}^{n_k} \varphi_\nu^{ij}-\tau_\nu  \mid\mathcal{D}_p \right] + \sum_{q=1}^{n_k} \mathbf{E}\left[ \frac{1}{m_kn_k}\sum_{i=1}^{m_k}\sum_{j=1}^{n_k} \varphi_\nu^{ij}-\tau_\nu  \mid \mathcal{D}'_q \right]\\
        & = \frac{1}{m_kn_k} \sum_{p=1}^{m_k}\sum_{i=1}^{m_k}\sum_{j=1}^{n_k} \mathbf{E}\left[\varphi_\nu^{ij}-\tau_\nu \mid\mathcal{D}_p\right] + \frac{1}{m_kn_k} \sum_{q=1}^{n_k} \sum_{i=1}^{m_k} \sum_{j=1}^{n_k} \mathbf{E} \left[\varphi_\nu^{ij} -\tau_\nu \mid \mathcal{D}'_q\right]\\
        :&= \mathbb{P + Q}.        
    \end{aligned}
\end{equation}
Firstly, let's explore the term $\mathbb{P}$ further. For given $p$,
\begin{itemize}
    \item  if $i\neq p,$ then $\mathbf{E}[\varphi_{\nu}^{ij}-\tau_\nu \mid\mathcal{D}_p]= \mathbf{E}[\varphi_{\nu}^{ij}-\tau_\nu] = 0$.
    \item  if $i = p,$ denote $\psi_\nu^{10}(\mathcal{D}_p):= \mathbf{E}[\varphi_{\nu}^{pj} \mid\mathcal{D}_p], $ which is a function of $\mathcal{D}_p,$ and we have $\mathbf{E} [\psi^{1,0}_\nu(\mathcal{D}_p)] =\tau_\nu$. To be noted, notation $\psi^\nu_{1,0}$ means that only the element of the treatment arm of $\varphi^{pj}$ is given with kernel function $\varphi_\nu$.
\end{itemize}
Similarly, we can ignore terms in $\mathbb{Q}$ when $j\ne q$, since their values are zero. Denote $\psi_\nu^{01}(\mathcal{D}'_q):=\mathbf{E}[\varphi_{\nu}^{iq} \mid \mathcal{D}'_q], $ which is a function of $\mathcal{D}'_q,$ and we have $\mathbf{E}[\psi_\nu^{01}(\mathcal{D}'_q)]=\tau_\nu.$  Notation $\psi_\nu^{01}$ means that only element of control arm of $\varphi^{iq}$ is given.
Thus the projection $\hat{U}_{\nu k}$ can be written as 
\begin{equation}\label{eq: Uk_Proj}
    \begin{aligned}
    \hat{U}_{\nu k}= \frac{1}{m_k}\sum_{i=1}^{m_k}\big\{\psi_\nu^{10}(\mathcal{D}_p)- \mathbf{E}\psi_\nu^{10}(\mathcal{D}_p)\big\} + \frac{1}{n_k}\sum_{j=1}^{n_k}\big\{\psi_\nu^{01}(\mathcal{D}'_q)- \mathbf{E}\psi_\nu^{01}(\mathcal{D}'_q)\big\} 
    \end{aligned}
\end{equation}

Furthermore, based on Theorem 12.6 in \cite{vander:98}, the variance of projection of $\hat{U}_{\nu k}$
\begin{equation}
    \begin{aligned}
    \mathbf{V}(\hat{U}_{\nu k}) 
    &= \frac{1}{m_k}\mathbf{V}\psi_\nu^{10}(\mathcal{D}) + \frac{1}{n_k} \mathbf{V}\psi_\nu^{01}(\mathcal{D}') = \frac{1}{m_k}\xi_{1,0}^{\nu \nu} + \frac{1}{n_k}\xi_{0,1}^{\nu \nu}. 
    \end{aligned}   
\end{equation}

Without loss of generality, we first show the asymptotic normality for the case when there are only 2 monitoring times in sequential design. As $m_1, n_1, m_2-m_1, n_2-n_1$ all go to $\infty,$ we have the following asymptotic normality results by multivariate Central limit theorem:
\[
\mathbf{S}_{m_1} := \frac{1}{\sqrt{m_1}}\sum_{i=1}^{m_1}
\begin{pmatrix}
    \psi_w^{10}(\mathcal{D}_p)- \mathbf{E}\psi_w^{10}(\mathcal{D}_p)\\
    \psi_l^{10}(\mathcal{D}_p)- \mathbf{E}\psi_l^{10}(\mathcal{D}_p)
\end{pmatrix}
\indist \mathcal{N}
\left(0, 
\begin{pmatrix}
    \xi^{10}_{ww} & \xi^{10}_{wl}\\
    \xi^{10}_{wl} & \xi^{10}_{ll}
\end{pmatrix}\right) := \mathcal{N}\left(\mathbf{0},\mathbf{\Sigma}_1\right),
\]
\[
\mathbf{S}_{m_2-m_1} := \frac{1}{\sqrt{m_2-m_1}}\sum_{i=m_1+1}^{m_2}
\begin{pmatrix}
    \psi_w^{10}(\mathcal{D}_p)- \mathbf{E}\psi_w^{10}(\mathcal{D}_p)\\
    \psi_l^{10}(\mathcal{D}_p)- \mathbf{E}\psi_l^{10}(\mathcal{D}_p)
\end{pmatrix}
\indist  \mathcal{N}\left(\mathbf{0},\mathbf{\Sigma}_1\right),
\]
\[
\mathbf{S}_{n_1} 
:= \frac{1}{\sqrt{n_1}}\sum_{j=1}^{n_1}
\begin{pmatrix}
    \psi_w^{01}(\mathcal{D}'_q)- \mathbf{E}\psi_w^{01}(\mathcal{D}'_q)\\
    \psi_l^{01}(\mathcal{D}'_q)- \mathbf{E}\psi_l^{01}(\mathcal{D}'_q)
\end{pmatrix}
\indist \mathcal{N}
\left(0, 
\begin{pmatrix}
    \xi^{01}_{ww} & \xi^{01}_{wl}\\
    \xi^{01}_{wl} & \xi^{01}_{ll}
\end{pmatrix}
\right) := \mathcal{N}\left(\mathbf{0},\mathbf{\Sigma}_2\right) ,
\]
and
\[
\mathbf{S}_{n_2-n_1} := \frac{1}{\sqrt{n_2-n_1}}\sum_{j=n_1+1}^{n_2}
\begin{pmatrix}
    \psi_w^{01}(\mathcal{D}'_q)- \mathbf{E}\psi_w^{01}(\mathcal{D}'_q)\\
    \psi_l^{01}(\mathcal{D}'_q)- \mathbf{E}\psi_l^{01}(\mathcal{D}'_q)
\end{pmatrix}
\indist \mathcal{N}\left(\mathbf{0},\mathbf{\Sigma}_2\right) 
\]

It is obviously that random vectors  $\mathbf{S}_{m_1}, \mathbf{S}_{m_2-m_1}, \mathbf{S}_{n_1},$ and $ \mathbf{S}_{n_2-n_1}$ are mutually independent. Then for arbitrary constant vector $\mathbf{c}_1, \mathbf{c}_2, \mathbf{c}_3$ and $\mathbf{c}_4,$
\begin{align*}
    & \Pr(\mathbf{S}_{m_1} < \mathbf{c}_1, \mathbf{S}_{m_2-m_1} < \mathbf{c}_2, \mathbf{S}_{n_1} < \mathbf{c}_3, \mathbf{S}_{n_2-n_1} < \mathbf{c}_4) \\
    =& \Pr(\mathbf{S}_{m_1} < \mathbf{c}_1) \Pr(\mathbf{S}_{m_2-m_1} < \mathbf{c}_2) \Pr(\mathbf{S}_{n_1} < \mathbf{c}_3) \Pr(\mathbf{S}_{n_2-n_1} < \mathbf{c}_4) \\
\rightarrow &  \Pr(\mathbf{Z}_1 < \mathbf{c}_1) \Pr(\mathbf{Z}_2 < \mathbf{c}_2) \Pr(\mathbf{Z}_3 < \mathbf{c}_3) \Pr(\mathbf{Z}_4 < \mathbf{c}_4) 
= \Pr(\mathbf{Z}_1 < \mathbf{c}_1, \mathbf{Z}_2 < \mathbf{c}_2, \mathbf{Z}_3 < \mathbf{c}_3, \mathbf{Z}_4 < \mathbf{c}_4),
\end{align*}
where $\mathbf{Z}_1, \mathbf{Z}_2, \mathbf{Z}_3$ and $\mathbf{Z}_4$ are mutually  independent normal 
distribution with $\mathbf{Z}_1, \mathbf{Z}_2 \sim  \mathcal{N} \left(\mathbf{0},\mathbf{\Sigma}_1\right) $ and $\mathbf{Z}_3, \mathbf{Z}_4 \sim\mathcal{N}\left(\mathbf{0},\mathbf{\Sigma}_2\right) 
.$ Thus, the joint distribution $\left(\mathbf{Z}_1,\mathbf{Z}_2,\mathbf{Z}_3,\mathbf{Z}_4\right)^\top \sim \mathcal{N} \left(
\mathbf{0}_{8\times1}, \mathbf{\Sigma}_{8\times8}^0\right),$
where 
$
\mathbf{\Sigma_{8\times8}^0}= 
\begin{pmatrix}
\mathbf{\Sigma}_1 & \mathbf{0} & \mathbf{0} & \mathbf{0} \\
\mathbf{0} & \mathbf{\Sigma}_1 & \mathbf{0} & \mathbf{0} \\
\mathbf{0} & \mathbf{0} & \mathbf{\Sigma}_2 & \mathbf{0} \\
\mathbf{0} & \mathbf{0} & \mathbf{0} & \mathbf{\Sigma}_2
\end{pmatrix}.
$ Therefore, as  $m_1, n_1, m_2-m_1, n_2-n_1$ all goes to $\infty,$ \begin{equation}
    \left(\mathbf{S}_{m_1}, \mathbf{S}_{m_2-m_1}, \mathbf{S}_{n_1}, \mathbf{S}_{n_2-n_1}\right)^\top \indist \mathcal{N} \left(
    \mathbf{0}_{4\times1}, \mathbf{\Sigma}_{8\times8}^0\right). 
\end{equation}

Then, it can be shown that
\begin{equation}\label{Eq: UProjCov}
   \left(\frac{\hat{U}_{w1}}{\sqrt{\mathbf{V}\left(\hat{U}_{w1}\right)}},\frac{\hat{U}_{l1}}{\sqrt{\mathbf{V}\left(\hat{U}_{l1}\right)}}, \frac{\hat{U}_{w2}}{\sqrt{\mathbf{V}\left(\hat{U}_{w2}\right)}},\frac{\hat{U}_{l2}}{\sqrt{\mathbf{V}\left(\hat{U}_{l2}\right)}} \right)^\top = \mathbf{M} \cdot \left(\mathbf{S}_{m_1}, \mathbf{S}_{m_2-m_1}, \mathbf{S}_{n_1}, \mathbf{S}_{n_2-n_1}\right)^T, 
\end{equation}
where
$$\mathbf{M} =
\begin{pmatrix}
\frac{1/\sqrt{m_1}}{\sqrt{\mathbf{V}(\hat{U}_{w1})}} & 0 & 0 & 0 & \frac{1/\sqrt{n_1}} {\sqrt{\mathbf{V}(\hat{U}_{w1})}} & 0 & 0 & 0  \\
0 & \frac{1/\sqrt{m_1}}{\sqrt{\mathbf{V}(\hat{U}_{l1})}}  & 0 & 0 & 0   & \frac{1/\sqrt{n_1}} {\sqrt{\mathbf{V}(\hat{U}_{l1})}}  & 0 & 0 \\
\frac{\sqrt{m_1}/m_2}{\sqrt{\mathbf{V}(\hat{U}_{w2})}} &  0 & \frac{\sqrt{m_2-m_1}/m_2}{\sqrt{\mathbf{V}(\hat{U}_{w2})}} & 0 & \frac{\sqrt{n_1}/n_2}{\sqrt{\mathbf{V}( \hat{U}_{w2})}} & 0 &\frac{\sqrt{n_2-n_1}/n_2}{\sqrt{\mathbf{V}(\hat{U}_{w2})}} & 0 \\
0 & \frac{\sqrt{m_1}/m_2}{\sqrt{\mathbf{V}(\hat{U}_{l2})}} & 0 & \frac{\sqrt{m_2-m_1}/m_2}{\sqrt{\mathbf{V}(\hat{U}_{l2})}} & 0 & \frac{\sqrt{n_1}/n_2}{\sqrt{\mathbf{V}( \hat{U}_{l2})}} & 0 &\frac{\sqrt{n_2-n_1}/n_2}{\sqrt{\mathbf{V}(\hat{U}_{l2})}} 
\end{pmatrix}.
$$
Owing to Slutsky's Theorem as it applies to matrices and vectors, and considering that a linear transformation of a normal distribution remains normal in the limiting distribution, it follows that

\begin{equation}\label{eq: tworawUProj}
   \left(\frac{\hat{U}_{w1}}{\sqrt{\mathbf{V}\left(\hat{U}_{w1}\right)}},\frac{\hat{U}_{l1}}{\sqrt{\mathbf{V}\left(\hat{U}_{l1}\right)}}, \frac{\hat{U}_{w2}}{\sqrt{\mathbf{V}\left(\hat{U}_{w2}\right)}},\frac{\hat{U}_{l2}}{\sqrt{\mathbf{V}\left(\hat{U}_{l2}\right)}} \right)^\top \indist \mathcal{N}\left(\mathbf{0}_{4 \times 1},\mathbf{\Sigma}_{4 \times 4}\right), 
\end{equation}
where $\mathbf{\Sigma}_{4 \times 4} = (\lim{\mathbf{M}}) \mathbf{\Sigma}^0_{8 \times 8} (\lim{\mathbf{M}}^\top) = \lim{(\mathbf{M \Sigma}^0_{8 \times 8} \mathbf{M}}^\top)$ and the limit means taking the limit as $m_1, n_1, m_2-m_1, n_2-n_1$ all goes to $\infty.$ By formula (\ref{Eq: UProjCov}), variance-covariance matrix  \[\mathbf{Cov}   \left(\frac{\hat{U}_{w1}}{\sqrt{\mathbf{V}\left(\hat{U}_{w1}\right)}},\frac{\hat{U}_{l1}}{\sqrt{\mathbf{V}\left(\hat{U}_{l1}\right)}}, \frac{\hat{U}_{w2}}{\sqrt{\mathbf{V}\left(\hat{U}_{w2}\right)}},\frac{\hat{U}_{l2}}{\sqrt{\mathbf{V}\left(\hat{U}_{l2}\right)}} \right)^\top = \mathbf{M \Sigma}^0_{8 \times 8} \mathbf{M}^\top \rightarrow \mathbf{\Sigma}_{4 \times 4}.\] To be noted, applying block multiplication on calculating $\mathbf{M \Sigma}^0_{8 \times 8} \mathbf{M}^\top$ make calculations simpler. For instance, we can partition $4\times 8$ dimensional matrix $\mathbf{M}$ into, eight $2\times 2$ diagonal square matrices. Similarly, we also can partition $8\times 8$ dimensional matrix $\mathbf{\Sigma}_{8\times 8}^0$ in to 16 squares matrices with dimension of $2\times 2.$ 

We can extend our proof of 2 times to the joint asymptotic normality for $K$ sequential monitoring times 
\begin{equation}
    \left(\frac{\hat{U}_{w1}}{\mathbf{V}\left(\hat{U}_{w1}\right)}, \frac{\hat{U}_{l1}}{\mathbf{V}\left(\hat{U}_{l1}\right)}, \frac{\hat{U}_{w2}}{\mathbf{V}\left(\hat{U}_{w2}\right)}, \frac{\hat{U}_{l2}}{\mathbf{V}\left(\hat{U}_{l2}\right)}, \ldots, \frac{\hat{U}_{wK}}{\mathbf{V}\left(\hat{U}_{wK}\right)}, \frac{\hat{U}_{lK}}{\mathbf{V}\left(\hat{U}_{lK}\right)}    \right)^\top \indist
    \mathcal{N}\left(\mathbf{0},\mathbf{\Sigma}_Z\right).
\end{equation}
Generally, $\mathcal{N}\left(\mathbf{0},\mathbf{\Sigma}_Z\right)$ represents the $2K$-dimensional multivariate normal distribution with a zero mean vector $\mathbf{0}$ of size $2K\times 1$, and a $2K\times 2K$ symmetric variance-covariance matrix $\mathbf{\Sigma}_Z$ with the diagonal elements are 1. As for off-diagonal elements,  $\forall \nu_1, \nu_2 = w, l$  and for $1\le p\leq q\le K$, and we have 
\begin{align*}
    &\mathbf{Cov}\left[\frac{\hat{U}_{\nu_1 p}}{\sqrt{\mathbf{V}(\hat{U}_{\nu_1 p})}}, \frac{\hat{U}_{\nu_2 q}}{\sqrt{\mathbf{V}(\hat{U}_{\nu_2 q})}}\right] 
    = \frac{m_q^{-1}\xi^{10}_{\nu_1\nu_2} + n_q^{-1}\xi^{01}_{\nu_1\nu_2}}{\sqrt{ \left(m_p^{-1}\xi^{10}_{\nu_1\nu_1} + n_p^{-1}\xi^{01}_{\nu_1\nu_1}\right)\times \left(m_q^{-1}\xi^{10}_{\nu_2\nu_2} + n_q^{-1}\xi^{01}_{\nu_2\nu_2}\right)}}\\
\rightarrow & \frac{\left( \sum_{k=1}^q\eta_k\right)^{-1} \xi^{10}_{\nu_1\nu_2} + \left(\sum_{k=1}^q\gamma_k\right)^{-1}\xi^{01}_{\nu_1\nu_2}} {\sqrt{ \left[\left(\sum_{k=1}^p\eta_k\right)^{-1}\xi^{10}_{\nu_1\nu_1} + \left(\sum_{k=1}^p\gamma_k\right)^{-1}\xi^{01}_{\nu_1\nu_1}\right] \times \left[\left(\sum_{k=1}^q\eta_k\right)^{-1}\xi^{10}_{\nu_2\nu_2} + \left(\sum_{k=1}^q\gamma_k\right)^{-1}\xi^{01}_{\nu_2\nu_2}\right]}}.
\end{align*}
as $m_k-m_{k-1}\to \infty$ and $n_k-n_{k-1}\to\infty$. 
$ \hfill\square$

\section{ Proof of Theorem 1}\label{app: Theorem1}

$\hat{U}_{\nu k}$ is the projection of $U_{\nu k}-\tau_\nu$ given by formula (\ref{Eq: RawUproj}) for $\nu = w,l, k=1,\ldots, K$. We have shown that the form of $\hat{U}_{\nu k}$  with variance in formula (\ref{eq: varRawProjU}). It is true that given Condition C1, as $n_{k}-n_{k-1}\rightarrow \infty, m_{k}-m_{k-1}\rightarrow \infty \text{ for } k = 1,\ldots, K, {\mathbf{V}(U_{\nu k})}/{\mathbf{V}(\hat{U}_{\nu k})}\rightarrow 1, \nu = w,l.$ Specifically, by Theorem 11.2 from van der Vaart (1998) \cite{vander:98},
\[
\frac{U_{\nu k} - \tau_\nu}{\sqrt{\mathbf{V}( U_{\nu k}})} - \frac{\hat{U}_{\nu k}}{\sqrt{\mathbf{V}(\hat{U}_{\nu k}})} \xrightarrow{P} 0, 
\]
for $ \nu = w,l; k = 1, 2, \ldots, K.$ This indicates that the difference between the normalized forms of \( \Delta U_k \) and \( \Delta \hat{U}_k \) converges to zero in probability. Additionally, it is evident that: $
\frac{\sqrt{\mathbf{V}(U_{\nu k})}}{\sqrt{\mathbf{V}( \hat{U}_{\nu k}})} \to 1,\text{ for } \nu = w,l; k = 1, 2, \ldots, K.
$

These findings, combined with Lemma~\ref{lem: RawProj}, demonstrate that \( \frac{\hat{U}_{\nu k}}{\sqrt{\mathbf{V}(\hat{U}_{\nu k})}} \) converges in distribution to a standard normal distribution, allow us to further deduce:
\begin{equation}
    \begin{aligned}
    \frac{U_{\nu k}- \tau_\nu - \hat{U}_{\nu k}}{\sqrt{\mathbf{V}(\hat{U}_{\nu k})}} &= \left( \frac{ U_{\nu k} - \tau_\nu}{\sqrt{\mathbf{V}( U_{\nu k})}} - \frac{ \hat{U}_{\nu k}}{\sqrt{\mathbf{V}(\hat{U}_{\nu k})}} \right) \frac{\sqrt{\mathbf{V}( U_{\nu k})}}{\sqrt{\mathbf{V}(\hat{U}_{\nu k})}} + \frac{ \hat{U}_{\nu k}}{\sqrt{\mathbf{V}( \hat{U}_{\nu k})}} \left( \frac{\sqrt{\mathbf{V}( U_{\nu k})}}{\sqrt{\mathbf{V}( \hat{U}_{\nu k})}} - 1 \right) \\
    &\xrightarrow{P} 0,
    \end{aligned}
\end{equation}
for \( \nu = w,l; k = 1, 2, \ldots, K \). This convergence in probability extends jointly to all estimators, leading us to conclude:

\begin{align}
 \left( \frac{ U_{w1} - \tau_{w} - \hat{U}_{w1}}{\sqrt{\mathbf{V} (\hat{U}_{w1})}},\frac{ U_{l1} - \tau_{l} - \hat{U}_{l1}}{\sqrt{\mathbf{V} (\hat{U}_{l1})}}, \ldots, \frac{ U_{wK} - \tau_{w} - \hat{U}_{wK}}{\sqrt{\mathbf{V} (\hat{U}_{wK})}}, \frac{ U_{lK} - \tau_{l} - \hat{U}_{lK}}{\sqrt{\mathbf{V} (\hat{U}_{lK})}} \right)^\top \xrightarrow{P} \mathbf{0}_{2K\times1}   ,
\end{align}
 as \( n_k - n_{k-1} \to \infty,  m_{k}-m_{k-1}\rightarrow \infty \), for \( k = 1, 2, \ldots, K \). Building upon these results, we apply Slutsky's Theorem on lemma \ref{lem: RawProj}. Given that \( \frac{\sqrt{\mathbf{V}(U_{\nu k})}}{\sqrt{\mathbf{V}( \hat{U}_{\nu k})}} \to 1 \) for each \( k \) and $\nu$, we establish the asymptotic normality of our estimators:
\begin{align}
\left( \frac{U_{w1} - \tau_w}{\sqrt{\mathbf{V}({U}_{w1})}}, \frac{U_{l1} - \tau_l}{\sqrt{\mathbf{V}({U}_{l1})}}, \ldots, \frac{U_{wK} - \tau_{w}}{\sqrt{\mathbf{V}({U}_{wK})}},  
\frac{ U_{lK} - \tau_{l}}{\sqrt{\mathbf{V}({U}_{lK})}}  \right)^\top \indist \mathcal{N}\left(\mathbf{0_{2K\times 1}}, \mathbf{\Sigma_{2K\times2K}}\right),
\end{align}
as \( n_k - n_{k-1} \to \infty \), for \( k = 1, 2, \ldots, K \). This completes our proof for the asymptotic normality of the estimators.

\section{Proof of Proposition 1}

\subsection{Proof of (1)}
Consider two test statistics $D_k$ and $D_s$ computed from $m_k$ and $m_s$ subjects in the treatment group and $n_k$ and $n_s$ subjects in the control group at the two time points $k$ and $s$, with $1\le k<s \le K$.
$$
\mathbf{Cov}\left(D_k,D_s\right)
=\mathbf{Cov}\left(\frac{\Delta U_k-\tau}{\sqrt{\mathbf{V} (\Delta U_k)}}, \frac{\Delta U_s-\tau}{\sqrt{\mathbf{V} (\Delta U_s)}}\right)=\frac{\mathbf{Cov}\left(\Delta U_k, \Delta U_s\right)}{\sqrt{\mathbf{V} (\Delta U_k)}\sqrt{\mathbf{V} (\Delta U_s)}}$$
If $\mathrm{Cov\left(\Delta U_k, \Delta U_s\right)}=\mathbf{V} (\Delta U_s)$, then $\mathbf{Cov}\left(D_k,D_s\right) = \sqrt{\mathbf{V} (\Delta U_s)/\mathbf{V} (\Delta U_k)}$ and this demonstrate that the independent increment assumption is met \cite{kim2020independent}.

\begin{equation}
    \begin{aligned}
    &\mathbf{Cov}\left(\Delta U_k, \Delta U_s\right)
     \\
    =&\mathbf{Cov} \left(\frac{1}{m_kn_k}\sum_{i=1}^{m_k}\sum_{j=1}^{n_k}\left( \varphi_w^{ij}-\varphi_l^{ij}\right),
    \frac{1}{m_sn_s} \sum_{p=1}^{m_s} \sum_{q=1}^{n_s} \left(\varphi_w^{pq}-\varphi_l^{pq}\right)\right) \\
    =&\frac{1}{m_kn_k}\frac{1}{m_sn_s}\Bigg[\sum_{i=1}^{m_k}\sum_{j=1}^{n_k}\mathbf{Cov}\left(\varphi_w^{ij}, \sum_{p=1}^{m_s} \sum_{q=1}^{n_s}\left(\varphi_w^{pq}-\varphi_l^{pq}\right)\right) -\sum_{i=1}^{m_k}\sum_{j=1}^{n_k}\mathbf{Cov}\left(\varphi_l^{ij}, \sum_{p=1}^{m_s} \sum_{q=1}^{n_s}\left(\varphi_w^{pq}-\varphi_l^{pq}\right)\right) \Bigg]\\
    = &\frac{1}{m_kn_k}\frac{1}{m_sn_s}\sum_{i=1}^{m_k}\sum_{j=1}^{n_k}\Bigg[ \mathbf{V} (\varphi_w^{ij}) + (n_s-1)\xi_{ww}^{10} + (m_s-1)\xi_{ww}^{01} - (n_s-1)\xi_{wl}^{10} -(m_s-1)\xi_{wl}^{01}-\mathbf{Cov}\left(\varphi_w^{ij}, \varphi_l^{ij}\right) \\
    &\quad \quad  \quad- \mathbf{Cov}\left(\varphi_l^{ij}, \varphi_w^{ij}\right) - (n_s-1)\xi^{10}_{lw} - (m_s-1)\xi^{01}_{lw}  + \mathbf{V} (\varphi_l^{ij}) + (n_s-1)\xi_{ll}^{10} + (m_s-1)\xi_{ll}^{01}\Bigg]\\
    =&\frac{1}{m_sn_s}\Bigg[ \mathbf{V} (\varphi_w^{ij}) + (n_s-1)\xi_{ww}^{10} + (m_s-1)\xi_{ww}^{01} - (n_s-1)\xi_{wl}^{10} -(m_s-1)\xi_{wl}^{01}-\mathbf{Cov}\left(\varphi_w^{ij}, \varphi_l^{ij}\right) \\
    &\quad \quad - \mathbf{Cov}\left(\varphi_l^{ij}, \varphi_w^{ij}\right) - (n_s-1)\xi^{10}_{lw} - (m_s-1)\xi^{01}_{lw}  + \mathbf{V} (\varphi_l^{ij}) + (n_s-1)\xi_{ll}^{10} + (m_s-1)\xi_{ll}^{01}\Bigg]\\
    =& \mathbf{Cov}\left(U_w^s, U_w^s\right) - \mathbf{Cov}\left(U_w^s, U_l^s\right) - \mathbf{Cov}\left(U_l^s, U_w^s\right) + \mathbf{Cov}\left(U_l^s, U_l^s\right)\\
    = & \mathbf{V}(\Delta U_s)
    \end{aligned}    
\end{equation}
Thus, $\mathbf{Cov}(D_k,D_s) = \sqrt{\mathbf{V}(\Delta U_s)/\mathbf{V} (\Delta U_k)}$.

\subsection{Proof of (2)}
According to Theorem 1 in the main manuscript, sequential win statistics 
$$\mathbf{Z}:=(Z_{w1}, Z_{l1}, \ldots, Z_{wK}, Z_{lK})^\top \indist\mathcal{N} \left(\mathbf{0},\mathbf{\Sigma}_Z\right),$$ where $Z_{\nu k}=\frac{U_{\nu k}-\tau_\nu}{\sqrt{\mathbf{V}(U_{\nu k})}}$ for $\nu = w, l; k=1,\ldots K,$ and $\mathbf{\Sigma}_Z$ is a $2K\times2K$ covariance-variance matrix.
Considering the form of IPCW-adjusted sequential Net Benefit statistics $\mathbf{D}=(D_1, D_2,\ldots, D_K)^\top,$ where $D_k=\frac{\Delta U_k-\Delta\tau}{\sqrt{\mathbf{V}(\Delta U_k)}}$, we construct $K\times 2K$ dimension matrix $\mathbf{A}$, such that \( \mathbf{D} = \mathbf{A} \cdot \mathbf{Z} \).
\begin{equation}
    \begin{aligned}
    \mathbf{A}&= \left(\begin{array}{ccccccc}
     \frac{\sqrt{\mathbf{V}(U_{w1})}}{\sqrt{\mathbf{V}(\Delta U_1)}}&  - \frac{\sqrt{\mathbf{V}(U_{l1})}}{\sqrt{\mathbf{V}(\Delta U_1)}} & & & & &\\
          &  &\frac{\sqrt{\mathbf{V}(U_{w2})}}{\sqrt{\mathbf{V}(\Delta U_2)}}&  - \frac{\sqrt{\mathbf{V}(U_{l2})}}{\sqrt{\mathbf{V}(\Delta U_2)}} & & &\\
     &  & & & \cdots & & \\
               &  & & & &\frac{\sqrt{\mathbf{V}(U_{wK})}}{\sqrt{\mathbf{V}(\Delta U_K)}} &  - \frac{\sqrt{\mathbf{V}(U_{lK})}}{\sqrt{\mathbf{V}(\Delta U_K)}}\\
\end{array}\right) \\
:&=
\left(
\begin{array}{cccc}
     A_1&    & &\\
     & A_2 & &  \\  
    & & \cdots & \\
    & & & A_K
\end{array}
\right).
    \end{aligned}
\end{equation}

The block matrix approach is applied to $\mathbf{A}$, where each block $\mathbf{A}_k$ is defined as 
\[ 
\mathbf{A}_k = \left( \frac{\sqrt{\mathbf{V}(U_{wk})}}{\sqrt{\mathbf{V}(\Delta U_k)}}, -\frac{\sqrt{\mathbf{V}(U_{lk})}}{\sqrt{\mathbf{V}(\Delta U_k)}} \right) \quad \text{for} \quad k = 1, \ldots, K.
\]

Similarly, applying the block matrix approach to $\mathbf{\Sigma}_{Z}$, it is partitioned into $K^2$ $2\times2$ matrices. Each block can be written as 
$\mathbf{\Sigma}_Z = \left[\mathbf{B}_{pq}\right]_{K \times K}$, 
where each block $\mathbf{B}_{pq}$ is given by
\[ 
\mathbf{B}_{pq} = \begin{pmatrix}
    \mathbf{Cov}(Z_{wp}, Z_{wq}) & \mathbf{Cov}(Z_{wp}, Z_{lq}) \\
    \mathbf{Cov}(Z_{lp}, Z_{wq}) & \mathbf{Cov}(Z_{lp}, Z_{lq})
\end{pmatrix}.
\]
Denote $\mathbf{\Sigma}_D = \lim \mathbf{A}\mathbf{\Sigma}_Z\mathbf{A}^\top,$ as $m_k-m_{k-1}\to \infty$ and $n_k-n_{k-1}\to\infty$ for $k=1,2,\cdots,K$. 
$ \mathbf{\Sigma}_D = \left[D_{pq}\right]_{K \times K}, $ where 
\begin{align*}
    D_{pq} 
    =&\lim \mathbf{A}_p\mathbf{B}_{pq}\mathbf{A}_q^\top \\
    =&\lim \left[\frac{\sqrt{\mathbf{V}(U_{wp})}}{\sqrt{\mathbf{V}(\Delta U_p)}} \times\mathbf{Cov}(Z_{wp}, Z_{wq})\times \frac{\sqrt{\mathbf{V}(U_{wq})}}{\sqrt{\mathbf{V}(\Delta U_q)}} + \frac{\sqrt{\mathbf{V}(U_{lp})}}{\sqrt{\mathbf{V}(\Delta U_p)}} \times\mathbf{Cov}(Z_{lp}, Z_{lq})\times \frac{\sqrt{\mathbf{V}(U_{lq})}}{\sqrt{\mathbf{V}(\Delta U_q)}}\right]  \\
    & - \lim \left[\frac{\sqrt{\mathbf{V}(U_{lp})}}{\sqrt{\mathbf{V}(\Delta U_p)}} \times\mathbf{Cov}(Z_{lp}, Z_{wq})\times \frac{\sqrt{\mathbf{V}(U_{wq})}}{\sqrt{\mathbf{V}(\Delta U_q)}} + \frac{\sqrt{\mathbf{V}(U_{wp})}}{\sqrt{\mathbf{V}(\Delta U_p)}} \times\mathbf{Cov}(Z_{wp}, Z_{lq})\times \frac{\sqrt{\mathbf{V}(U_{lq})}}{\sqrt{\mathbf{V}(\Delta U_q)}}\right] \\
    &:= \mathbb{R- S}.
\end{align*}
By independent increment, 
\begin{align*}
\mathbb{R} &=\lim \frac{\mathbf{V}(U_{wq})+\mathbf{V}(U_{lq})}{\sqrt{\mathbf{V}(\Delta U_p)\mathbf{V}(\Delta U_q)}}= \frac{({\sum_{k=1}^q\eta_k})^{-1}\times {(\xi^{10}_{ww} + \xi^{10}_{ll})} + ({\sum_{k=1}^q\gamma_k}) ^{-1}\times({\xi^{01}_{ww} + \xi^{01}_{ll}})}{\sqrt{\left[\frac{\xi^{10}_{ww} + \xi^{10}_{ll}-2\xi^{10}_{wl}}{\sum_{k=1}^p\eta_k}  + \frac{\xi^{01}_{ww} + \xi^{01}_{ll}-2\xi^{01}_{wl}}{\sum_{k=1}^p\gamma_k} \right]\times \left[\frac{\xi^{10}_{ww} + \xi^{10}_{ll}-2\xi^{10}_{wl}}{\sum_{k=1}^q\eta_k}  + \frac{\xi^{01}_{ww} + \xi^{01}_{ll}-2\xi^{01}_{wl}}{\sum_{k=1}^q\gamma_k} \right] }} .
\end{align*}
And, also we can simplify 
\begin{align*}
    \mathbb{S}= \frac{({\sum_{k=1}^q\eta_k})^{-1}\times {2\xi^{10}_{wl}} + ({\sum_{k=1}^q\gamma_k})^{-1}\times {2\xi^{01}_{wl}}}{\sqrt{\left[\frac{\xi^{10}_{ww} + \xi^{10}_{ll}- 2\xi^{10}_{wl}}{\sum_{k=1}^p\eta_k}  + \frac{\xi^{01}_{ww} + \xi^{01}_{ll}-2\xi^{01}_{wl}}{\sum_{k=1}^p\gamma_k} \right]\times \left[\frac{\xi^{10}_{ww} + \xi^{10}_{ll}-2\xi^{10}_{wl}}{\sum_{k=1}^q\eta_k}  + \frac{\xi^{01}_{ww} + \xi^{01}_{ll}-2\xi^{01}_{wl}}{\sum_{k=1}^q\gamma_k} \right] }}
\end{align*}
Thus, \[
D_{pq} = \mathbb{R - S} =\sqrt{\frac{\left(\sum_{k=1}^q\eta_k\right)^{-1} \left(\xi^{10}_{ww} + \xi^{10}_{ll}-2\xi^{10}_{wl}\right) + \left(\sum_{k=1}^q\gamma_k\right)^{-1}\left(\xi^{01}_{ww} + \xi^{01}_{ll}-2\xi^{01}_{wl}\right)} {\left(\sum_{k=1}^p\eta_k\right)^{-1} \left(\xi^{10}_{ww} + \xi^{10}_{ll}-2\xi^{10}_{wl}\right) + \left(\sum_{k=1}^p\gamma_k\right)^{-1} \left(\xi^{01}_{ww} + \xi^{01}_{ll}-2\xi^{01}_{wl}\right)}}.
\]

\section{Proof of Proposition 2}\label{app: PropURatio}

\subsection{Proof of part (1) of Proposition 2}
Similarly setting with \ref{App: Lemma1}, consider two test statistics $R_k$ and $R_s$ computed from $m_k$ and $m_s$ subjects in the treatment group and $n_k$ and $n_s$ subjects in the control group at the two time points $k$ and $s$, with $1\le k<s \le K$.
$$
\mathbf{Cov}\left(R_k, R_s\right)
=\mathbf{Cov}\left(\frac{\log{(\Theta U_k)}-\log{(\Theta\tau)}}{\sqrt{\mathbf{V} [\log{(\Theta  U_k)]}}}, \frac{\log{(\Theta U_s)}-\log{(\Theta\tau)}}{\sqrt{\mathbf{V} [\log{(\Theta  U_s)]}}}\right) =\frac{\mathbf{Cov}\left[\log{(\Theta U_k)}, \log{(\Theta U_s)}\right]}{\sqrt{\mathbf{V} [\log{(\Theta U_k)}]}\sqrt{\mathbf{V} [\log{(\Theta U_s)}]}}$$
If $\mathbf{Cov}\left[\log{(\Theta U_k)}, \log{(\Theta U_s)}\right] =\mathbf{V} [\log{(\Theta U_s)}]$, then $\mathbf{Cov}\left(R_k,R_s\right) = \sqrt{\mathbf{V} [\log{(\Theta  U_s)]}/\mathbf{V} [\log{(\Theta  U_k)]}}$ and this demonstrate that the independent increment assumption is met \cite{kim2020independent}. The below proof follows a similar idea from Bergemann \textit{et al.} (2023) \cite{bergemann4450923group}.

As is shown in the derivation of $\mathbf{V}\left[\log U_u\right]$ in Appendix \ref{App: VRatioU}, we have the below approximation by Taylor expansion for moments of functions of random variables $X_1, X_2$:
$$
{\mathbf{Cov}\left(\log{X_1},\log{X_2}\right)}\approx \log\left(\frac{\mathbf{Cov}(X_1,X_2)}{\mathbf{E}X_1\mathbf{E}X_2}+1\right).
$$
Given the above results, 
\begin{align*}
    \mathbf{Cov}\left[\log{(\Theta U_k)}, \log{(\Theta U_s)}\right]
    &= \mathbf{Cov}\left(\log{U_{wk}} - \log{U_{lk}}, \log{U_{ws}} - \log{U_{ls}}\right) \\
    &= \mathbf{Cov}\left(\log{U_{wk}}, \log{U_{ws}}\right) - \mathbf{Cov}\left(\log{U_{wk}}, \log{U_{ls}}\right) \\
    &\quad -\mathbf{Cov}\left(\log{U_{lk}}, \log{U_{ws}}\right) + \mathbf{Cov}\left(\log{U_{lk}}, \log{U_{ls}}\right) \\
    &\approx \log {\left( \frac{\mathbf{Cov}(U_{wk}, U_{ws})}{\tau_w \tau_w} +1\right)} - \log \left( \frac{\mathbf{Cov}(U_{wk}, U_{ls}) }{\tau_w \tau_l}+1 \right)  \\
    &\quad  - \log \left( \frac{\mathbf{Cov}(U_{lk}, U_{ws})}{\tau_l \tau_w}+1 \right) + \log \left( \frac{\mathbf{Cov}(U_{lk}, U_{ls}) }{\tau_l \tau_l} +1\right) \tag{By Taylor Expansion Results} \\
     &= \log \left( \frac{\mathbf{Cov}(U_{ws}, U_{ws}) }{\tau_w \tau_w} +1\right) - \log \left( \frac{\mathbf{Cov}(U_{ws}, U_{ls})}{\tau_w \tau_l} +1 \right)   \\
     &\quad- \log \left( \frac{\mathbf{Cov}(U_{ls}, U_{ws})}{\tau_l \tau_2} +1\right) + \log \left( \frac{\mathbf{Cov}(U_{ls}, U_{ls}) }{\tau_l \tau_l} +1 \right) \tag{By Results from Lemma 1}
\end{align*}

Plugin the results above back to $ \mathbf{Cov}\left[\log{(\Theta U_k)}, \log{(\Theta U_s)}\right]$, we obtain
\begin{align*}
        \mathbf{Cov}\left[\log{\Theta U_k}, \log{\Theta U_s}\right]
        &\approx \mathbf{Cov}[\log{U_{ws}}, \log{U_{ws}}] - \mathbf{Cov}[\log{U_{ws}}, \log{U_{ls}}] \\
        & \quad -\mathbf{Cov}[\log{U_{ls}}, \log{U_{ws}}] + \mathbf{Cov}[\log{U_{ls}}, \log{U_{ls}}] \\
        &= \mathbf{Cov}\left[\log{U_{ws}} - \log{U_{ls}}, \log{U_{ws}} - \log{U_{ls}}\right] \\
        &= \mathbf{V}  \left[ \log{(\Theta U_s)}\right]\\
\end{align*}
Thus, $\mathbf{Cov}(R_k,R_s) = \sqrt{\mathbf{V} [\log{(\Theta  U_s)]}/\mathbf{V} [\log{(\Theta  U_k)]}}$

\end{document}